\documentclass[aps,%
               prl,%
               reprint,%
               showpacs,%
               superscriptaddress,%
               longbibliography,%
               nofootinbib%
               ]{revtex4-2}

\usepackage[utf8]{inputenc}
\usepackage[usenames,dvipsnames]{xcolor}
\usepackage{graphicx}
\usepackage{amsfonts}
\usepackage{amssymb}
\usepackage{amsmath}
\usepackage{mathtools}
\usepackage{bm}
\usepackage{dsfont}
\usepackage{braket}
\usepackage{tikz}
\usepackage{txfonts}

\usepackage[pdfusetitle,%
            bookmarks=true,%
            colorlinks,%
            linkcolor=blue,%
            citecolor=blue,%
            urlcolor=blue%
            ]{hyperref}

\newcommand{\eq}[1]{Eq.\thinspace(\ref{#1})}
\newcommand{\eqs}[2]{Eqs.\thinspace(\ref{#1},\ref{#2})}

\newcommand{\fig}[1]{Fig.\thinspace{}\ref{#1}}

\newcommand{\subfigref}[2]{\hyperref[fig:#1]{\ref*{fig:#1}(#2)}}

\newcommand{\TUM}{\affiliation{Technical University of Munich, TUM School of Natural Sciences, Physics Department, 85748 Garching, Germany}}
\newcommand{\MCQST}{\affiliation{Munich Center for Quantum Science and Technology (MCQST), Schellingstr. 4, 80799 M{\"u}nchen, Germany}}
\newcommand{\MIT}{\affiliation{Center for Theoretical Physics - a Leinweber Institute, Massachusetts Institute of Technology, Cambridge, MA 02139, USA}}

\begin{document}
\def\papertitle{{Skeleton of isometric Tensor Network States for Abelian String-Net Models}}
\title{\papertitle}

\author{Julian Boesl} \TUM \MCQST
\author{Yu-Jie Liu} \MIT
\author{Frank Pollmann}  \TUM \MCQST
\author{Michael Knap}  \TUM \MCQST

\date{\today}

\begin{abstract}
We construct parametrized isometric tensor network states---referred to as skeletons---that allow us to explore phases of abelian topological order and can be efficiently implemented on quantum processors. We obtain stable finite correlation length deformations of string-net fixed points, which are constructed both by conserving virtual symmetries of the tensor and by imposing local isometry constraints. They connect distinct topological phases via a shared critical point, thereby providing analytically tractable examples of phase transitions beyond anyon condensation. By mapping such classes of 2D tensor networks to 1D stochastic automata with local update rules, we show that expectation values of generalized Pauli strings of arbitrary weight can be efficiently computed using classical methods.
Therefore these skeletons not only serve as an organizing principle for abelian topological order but also provide a non-trivial testbed for quantum processors.
\end{abstract}

\maketitle
\textbf{\emph{Introduction.---}}Tensor network states (TNS) have emerged as an efficient tool to understand topological order~\cite{cirac_matrix2021, verstraete_criticality2006, schuch_peps2010, sahinoglu_characterizing2021, williamson_symmetry2017}. Besides their strength as variational states, a number of exact wavefunction paths built from finite tensors are known which interpolate between different fixed points. In one dimension, skeletons of matrix product states (MPS) have been identified which connect distinct symmetry-protected topological orders~\cite{wolf_quantum2006, jones_skeleton2021, camp_matrix2025}. Such a transition has been implemented on a quantum device using a sequential circuit deduced from the MPS~\cite{smith_crossing2022}. In higher dimensions, paths between topologically ordered and trivial phases can also be constructed~\cite{chen_tensor2010, xu_tensor2018, haller_quantum_2023}. However, these are more unwieldy than their 1D counterparts, as  general 2D TNS are less tractable~\cite{haegeman_shadows2015, iqbal_study2018, xu_characterization2021} and do not have an efficient preparation protocol with unitary circuits~\cite{schuch_computational_2007}. Yet, fixed points of phases often exhibit more favorable properties: For instance, the fixed points of string-net models are isometric TNS (isoTNS), which allow for an efficient contraction of the bare tensor network and can be realized by means of a linear-depth sequential circuit~\cite{zaletel_isometric2020, soejima_isometric2020, slattery_quantum_2021, wei_sequential2022,liu:2022}. Paths which keep these properties would thus serve as valuable tools to study topological phases and their transitions. Specific examples for such paths have already been introduced~\cite{liu_simulating2024, yu_dual2024, boesl_quantum2025}. However, it remains open how such deformations can be systematically constructed.

Here, we propose classes of exact continuous isoTNS obtained from abelian string-net models~\cite{levin_string2005, gu_tensor2009, buerschaper_explicit2009, hung_string2012, lan_topological2014, lin_generalizations2014, lin_generalized2021} which allow for analytical insights and precise numerical predictions. Leveraging the isometry condition on the tensors, we show that they belong to a set of quantum states which map to stochastic circuits, with abelian string-nets corresponding to update rules conserving parity-like quantities. This mapping ensures the presence of the topological order, as the gap of the associated circuit closes only when entries in the tensor are tuned to zero. By connecting different fixed points to such critical points, we find topological phase transitions beyond the anyon condensation paradigm~\cite{gu_tensor2008, vidal_lowenergy2009, dusuel_robustness2011, wu_phase2012, burnell_anyon2018, lin_anyon2024, cochran_visualizing2025}. This strategy allows us to map out the space of possible deformations as represented in \fig{fig:Skeletons}: Fixed points with the same local Hilbert space dimension form part of an interwoven network of stable directions with phase transitions at directly identifiable points; we call this network an \emph{isoTNS skeleton}. 
Furthermore, by utilizing that isoTNS skeletons map to stochastic circuits, we show that generalized Pauli strings of arbitrary weight can be evaluated efficiently via classical sampling methods. 
General operators, that are an exponential superposition of Pauli strings, can be measured classically only for sufficiently small support. Due to the relation between isoTNS and linear-depth circuits, these states also serve as excellent higher-dimensional testbeds for quantum devices, supporting finite correlation length and non-trivial many-body ordering, while still exhibiting a set of properties that admits efficient classical verification.

\begin{figure}
    \includegraphics[width=.98\columnwidth]{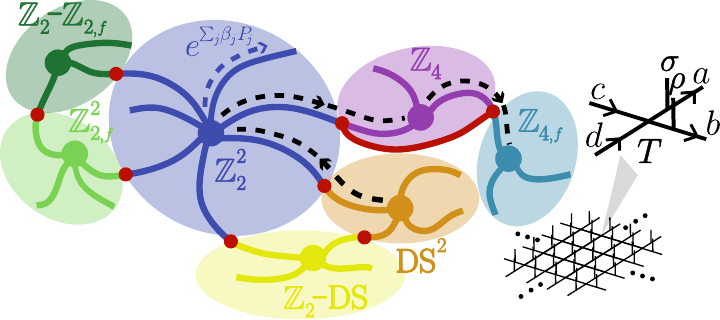}
    \caption{\label{fig:Skeletons}
        \textbf{Skeleton of isometric Tensor Network States.} Starting from abelian string-net models, we construct classes of isometric tensor network states within the topological phase of matter, connected by an imaginary time evolution. At some points (red dots), a critical state is reached from different topologically ordered phases, realizing a quantum phase transition. For local Hilbert space dimension $N = 4$ possible phases include pairs of toric code ($\mathbb{Z}_2$), pairs of double-semion (DS) states, or the $\mathbb{Z}_4$ toric code. The subscript $f$ denotes phases with non-trivial symmetry fractionalization under an anti-unitary symmetry. The dashed path is further investigated in \fig{fig:Path}. The isometric tensor network is illustrated on the right.
    }
\end{figure}

\textbf{\emph{String-Net Models.---}} A large number of gapped topologically ordered phases are represented by string-net models~\cite{levin_string2005, lin_generalizations2014, lin_generalized2021}. We consider abelian string-net models, which are fixed points for many abelian topological orders. We explicitly discuss the case of the underlying abelian group $G$ being a cyclic group $G = \mathbb{Z}_N$; other finite abelian groups can be treated equally as they are isomorphic to products of cyclic groups.

We consider a square lattice with a qudit of local Hilbert space dimension $N$ on each edge. For a given $N$, we introduce the clock operator $Z$ as $Z\ket{j} = e^{2\pi ij/N} \ket{j}$ and the shift operator $X$ as $X\ket{j} = \ket{(j + 1 )\;\text{mod} \; N}$. Abelian string-net states can be written graphically as
\begin{equation}
    \includegraphics[scale=0.425]{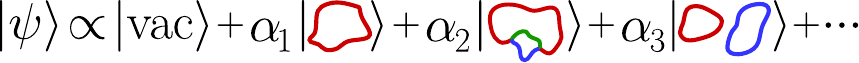};
    \label{eq:PsiDef}
\end{equation}
colors indicate string types. At the fixed points, $\vert \alpha_c\vert= 1$ for every configuration $c$: The state is an equal-weight superposition of all closed-loop configurations, for which a parent Hamiltonian of commuting local projectors exists~\cite{levin_string2005, perezgarcia_peps2007} (see SM~\cite{supp} Sec. A and C for a more thorough explanation).

String-net states can be expressed as tensor network states (TNS)~\cite{gu_tensor2009, buerschaper_explicit2009, shukla_boson2018}. Introducing a tensor $T_{abcd}^{\sigma\rho}$ at every vertex, with virtual legs $a,b,c,d$ of bond dimension $D$ and physical legs $\sigma,\rho$ on two edges, the state is obtained by contracting the virtual legs,
\begin{equation}
        \ket{\psi} = \sum_{\sigma_1, \cdots, \sigma_k} \text{tTr}\left( \left\{ T^{\sigma_1\sigma_2}, \cdots, T^{\sigma_{k-1}\sigma_k} \right\} \right) \ket{\sigma_1\cdots\sigma_k},
    \label{eq:TensorWF}
\end{equation}
where $\text{tTr}$ denotes this tensor contraction.

When the phases $\alpha_c$ depend on local properties as in the $\mathbb{Z}_N$ toric code, the state can be described by so-called \emph{single-line} TNS. The tensors lock physical and virtual legs and are decomposed as $T_{abcd}^{\sigma\rho} = \sum_{a^\prime b^\prime} \delta_{aa^\prime}^\sigma \delta_{bb^\prime}^\rho W_{(a^\prime b^\prime )(cd)}$, or graphically
\begin{equation}
    \includegraphics[scale=0.5]{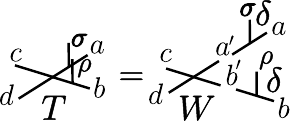},
    \label{eq:SingleLine} 
\end{equation}
where $W$ is an $N^2 \times N^2$ matrix and $\delta$ is a ``plumbing" tensor with $\delta_{ab}^\sigma = 1$ if $a = b = \sigma$ and zero otherwise. The bond dimension is equivalent to the physical dimension $D = N$.

A more general set which describes all abelian string-net models are \emph{double-line} TNS.
They allow for the non-local phase factors of ``non-toric code" string-net states~\cite{levin_braiding2012, zhang_nonhermitian2020, xu_tensor2018}. The tensor has bond dimension $D = N^2$ and can be decomposed into a $N^4 \times N^4$ $W$-matrix and tensors on the edges relating virtual and physical legs (see SM~\cite{supp} Sec. A for a precise definition of the TNS).
The TNS representation of every string-net fixed point preserves certain virtual symmetries that are associated with the underlying topological order. The virtual symmetries are implemented by operators that act solely within the total virtual Hilbert space, or more generally, virtual subspaces, spanned by the virtual legs of the tensor network (See SM \cite{supp}, Sec. B for further details). 

Single- and double-line representations of string-nets are isometric TNS (isoTNS), which form an expressive subclass of TNS with a contraction property similar to the canonical form of matrix product states; for single-line tensors, it is $\sum_{\sigma,\rho,c,d}T^{\sigma\rho}_{abcd}(T^*)^{\sigma\rho}_{a^\prime b^\prime cd} = \delta_{aa^\prime}\delta_{bb^\prime}$~\cite{zaletel_detecting_2014}. A related structure holds for double-line tensors as well. IsoTNS can be prepared using a linear-depth unitary circuit~\cite{slattery_quantum_2021, wei_sequential2022,liu:2022, liu_simulating2024}, making them accessible to current quantum devices~\cite{satzinger_realizing_2021, smith_simulating2025}. 

\textbf{\emph{Mapping to Stochastic Automata.---}} In general, a single-line state defined as in \eq{eq:SingleLine} is an isoTNS if its $W$-matrix is \emph{normalized}~\cite{liu_simulating2024} as
\begin{equation}
    \sum_{c,d} \vert W_{(ab)(cd)}\vert^2 = 1, \; \forall a,b.
    \label{eq:WNorm}
\end{equation}
This allows for direct identification of the $W$-matrix with the local update rule of a stochastic automaton: The value $\vert W_{(ab)(cd)}\vert^2$ expresses the probability $p(cd\vert ab)$ that two sites in the local states $a$ and $b$ change into states $c$ and $d$. As physical and virtual legs are equivalent for a single-line tensor, the two-dimensional quantum state is mapped to a stochastic automaton circuit acting on a one-dimensional state as shown in \fig{fig:Mapping}: Each tensor is an update step in a brickwork circuit, where space and time are identified with the diagonal directions of the original state. Starting from an open boundary, the quantum state is thus a superposition of all allowed trajectories weighed by the square root of their probability,  possibly with complex phases (see also~\cite{zhang_sequential2025}). 

\begin{figure}
    \includegraphics[width=.995\columnwidth]{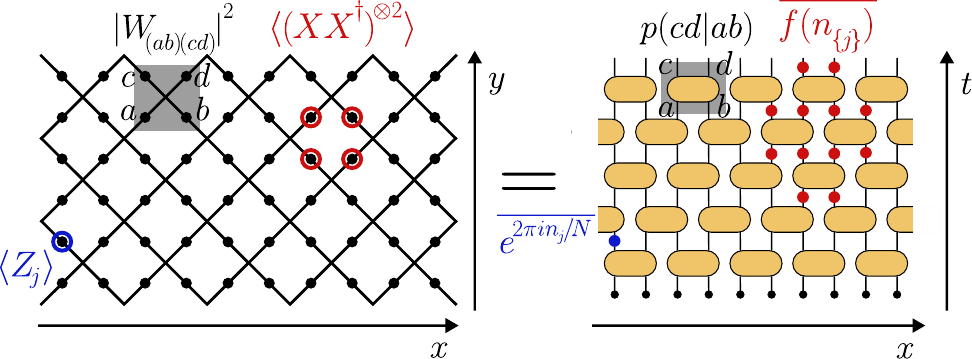}
    \caption{\label{fig:Mapping}
        \textbf{Mapping to stochastic automata.} The normalization property of the single-line $W$-matrix (left) allows us to identify it with the two-local update rule of a stochastic automaton (right). From this, a tensor network state with an open boundary can be interpreted as a superposition of all paths of this stochastic process, where one spatial direction of the quantum state becomes the time direction of the automaton. Diagonal operators such as $Z_j$ on site $j$ can be evaluated with classical Monte Carlo methods from the local state $n_j$ in the stochastic circuit. Generalized Pauli string operators such as loop operators $(XX^\dagger)^{\otimes2}$ (red) can be mapped to functions of the classical state at different points in time and space $f(n_{\{j\}})$, and thereby efficiently evaluated as well.
    }
\end{figure}

The state at the boundary of the tensor network may be any matrix-product state, including the simplest case of a product state. For product states, local diagonal operators can be directly evaluated using classical sampling methods. For example, the expectation value of the clock operator $Z(x,y)$ at a site given by coordinates $x$ and $y$ as shown in \fig{fig:Mapping} can be evaluated using the identification
$\langle Z(x,y) \rangle = \overline{ e^{2\pi i n_x(t  = y) /N }}$,
where $n_x(t)$ is the value of the automaton at site $x$ and time $t$ and the average on the right-hand side is over realizations of the stochastic dynamics, which are initialized in the distribution given by the boundary state. This identification extends to all generalized Pauli strings formed by $Z^a X^b, \forall a,b \in \{0,N-1\}$ and $a + b > 0$ (excluding the identity so that the number of non-identity operators sets the weight of the Pauli string). This follows from absorbing the action of any non-diagonal Pauli operators (such as $X$) into Pauli operators at the virtual legs using the plumbing tensor
\begin{equation}
    \includegraphics[scale=0.425]{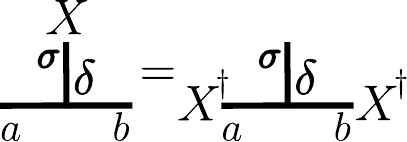}.
    \label{eq:main_Plumbing}
\end{equation}
The virtual Pauli operators modify the local $W$-matrices of the state, and thereby the coefficients of the wavefunction in the computational basis. Their action is equivalent to an operator diagonal in the computational basis; details of the procedure are presented in SM~\cite{supp} Sec. D. In particular, we show how for any Pauli string $O$ (of arbitrary weight) a diagonal operator $O_{\text{diag}}$, with support that is at most a constant multiple of that of the original operator $O$, can be explicitly constructed such that $\langle O \rangle = \langle O_{\text{diag}} \rangle$. Crucially, $O_{\text{diag}}$ can always be decomposed into a product of local diagonal operators $O_v$, each of which acts on the four qubits adjacent to the vertex $v$, enabling efficient evaluation via the classical automaton (up to any fixed additive error).
For example, a local loop operator of $(XX^\dagger)^{\otimes2}$ corresponds to a function of 12 space-time points in the classical automaton; see \fig{fig:Mapping}. We emphasize that this property holds for all normalized single-line tensor-network states, including single-line TNS in higher dimensions and on lattices with higher connectivity.

These identifications do not hold for generic tensor networks, as different virtual indices may correspond to the same local physical state. In general, it is neither possible to evaluate diagonal operators in terms of stochastic processes nor to map an operator to a corresponding diagonal operator. The situation is however much more favorable for double-line states which preserve the virtual symmetries of a string-net model. The string-net fixed point $\ket{\psi_{\text{fix}}} = U \ket{\mathbb{Z}_N}$ is connected to the $N$-level qudit toric code $\ket{\mathbb{Z}_N}$ by a global diagonal unitary $U$. At the same time, a deformed state $\ket{\psi} = \mathcal{T} \ket{\psi_{\text{fix}}}$ conserving all double-line symmetries is connected to its respective fixed point by a diagonal imaginary time evolution $\mathcal{T}$, which is constructed from the physical perturbations associated to the tensor deformation (see SM~\cite{supp} Sec. B). The expectation value of a diagonal operator $O_{\text{diag}}$ can be rewritten as
\begin{equation}
    \bra{\psi}O_{\text{diag}}\ket{\psi} = \bra{\mathbb{Z}_N}U^\dagger\mathcal{T}^\dagger O_{\text{diag}}\mathcal{T}U\ket{\mathbb{Z}_N} = \bra{\psi^\prime}O_{\text{diag}}\ket{\psi^\prime},
    \label{eq:DLtoSL}
\end{equation}
where $\ket{\psi^\prime} = \mathcal{T}\ket{\mathbb{Z}_N}$ is a deformed $\mathbb{Z}_N$ toric code and can thus be expressed with single-line tensors. Here, we have used that all operators $\mathcal{T}, O_{\text{diag}}, U$ are diagonal and therefore commute. If the original double-line tensor of $\ket{\psi}$ further is an isoTNS, its double-line $W$-matrix fulfills a normalization condition similar to \eq{eq:WNorm}. Then, the single-line $W$-matrix of $\ket{\psi^\prime}$ is also normalized and can be mapped to a stochastic circuit with an update rule which can be directly read off from the original double-line tensor (see SM~\cite{supp} Sec. D). At the same time, we can use the closed-loop constraint to again map generalized Pauli strings to diagonal operators with the same expectation value. These can be efficiently evaluated classically, see SM~\cite{supp} Sec. D. Therefore, for normalized double-line states which respect string-net virtual symmetries the same techniques are applicable as for normalized single-line states. For both classes of normalized TNS, generic operators with a large support $k$ are in general composed of $N^{2k}$ generalized Pauli strings, and thereby remain difficult to evaluate classically; operators with small support can still be evaluated by treating each Pauli string separately.

As both classes of states discussed in this section, normalized single-line states (including normalized single-line string-nets) and normalized double-line string-nets, are isoTNS by construction, they can serve as ideal benchmarking states for quantum processors. They occupy a sweet spot between classical and quantum accessibility and non-trivial features: They can be realized on a quantum device by an efficient quantum circuit, and allow at the same time for classical evaluation of a large class of physically relevant operators, while still being able to represent topological order at finite correlation length, as we are going to discuss in the next section. 

\textbf{\emph{IsoTNS Skeletons.---}} 
General deformations of the string-net fixed point tensors that preserve the virtual symmetries lead to local physical perturbations~\cite{shukla_boson2018}. Thus, topological order remains stable to such perturbations, provided they are weak enough. Determining up to which strength of the perturbation topological order is robust before undergoing a phase transition becomes a formidable task that typically needs to be solved numerically~\cite{iqbal_study2018, xu_characterization2021}. A simple example of such a transition which still allows for an exact solution is the toric code with string tension (generated by an imaginary time evolution with a field). This can be written as a single-line tensor which is not isometric away from the fixed point. Above a critical tension, the plaquette anyons condense and drive the state into a paramagnetic phase~\cite{chen_tensor2010, papanikoaou_topological2007, castelnovo_quantum2008}.

Our strategy is distinct: we want to find classes of states for which the presence of topological order is assured unless fine-tuned conditions are met. We achieve this by considering classes of tensors that preserve both (i) virtual symmetries and at the same time (ii) the isometry condition on the tensor. Property (i) is necessary for the robustness of topological order and property (ii) allows us to map such states to stochastic circuits, as discussed in the preceding section. Since the one-dimensional stochastic automata remain gapped when deforming their weights while keeping the symmetries of the fixed point of the circuit, the corresponding string-net fixed point will also remain in the same topologically ordered phase.

In the stochastic automaton, every virtual symmetry is expressed as conservation of parity (or a higher-modulo generalization) in the local updates. For instance, the MPO symmetry of a $\mathbb{Z}_N$ toric code enforces conservation of the quantity $P_{N,j} = \left(\sum_{x \in j} n_x \right) \;\text{mod}\; N$ in every gate $j$. Accordingly, any update rule which respects such a parity-like conservation law provides a deformed single/double-line tensor for each abelian string-net with the corresponding local branching constraint. Following this general strategy, these stochastic circuits provide us with classes of states with finite correlation length $\xi$, which belong to the same topological order as the respective fixed point.

These stochastic automata become gapless only when fine-tuned to higher-symmetry points, which determine critical end points of the topologically ordered phases. Accordingly, the critical isoTNS must have a higher virtual symmetry than the topological isoTNS it directly connects with; this happens only when some transition amplitudes/tensor entries $p(cd\vert ab) = \vert W_{(ab)(cd)}\vert^2$ vanish. We identify critical points whose enhanced symmetries include the virtual symmetries of multiple phases: At these points, a continuous transition between these phases occurs. With that we construct an interwoven web of parametrized paths that allow us to systematically connect distinct quantum phases, including both topologically ordered and trivial phases. We refer to this web as an \emph{isoTNS skeleton}; see Fig.~\ref{fig:Skeletons}.

We see that deformed isoTNS host continuous quantum phase transitions without anyon condensation. Critical points are  realized when the gap of the transfer matrix shared by the quantum state and the stochastic automaton closes. (It should be noted that this is the transfer matrix in the time-like direction $y$, i.e., it is rotated by 45 degrees compared to the unit vectors of the square lattice.) As an example, we construct a path which crosses multiple fixed points with the same local Hilbert space dimension $N = 4$; see \fig{fig:Path} and Sec. E of the SM~\cite{supp}. The path includes pairs of $\mathbb{Z}_2$ string-net layers as well as the intrinsic $\mathbb{Z}_4$ string-nets. In total, four distinct phases are explored: We start with a phase transition from a pair of double-semion models (DS) to a pair of $\mathbb{Z}_2$ toric codes, where the models differ only in the braiding statistics of their excitations (yellow to violet)~\cite{xu_tensor2018}. This transition needs a double-line description as the sign structure of the double-semion model is non-local. From the fixed point of two decoupled toric code layers, we pass a phase transition to the $\mathbb{Z}_4$ toric code (violet to purple). This is an example of a more general coupling of $k$-layer $\mathbb{Z}_N$ toric code states to a $\mathbb{Z}_{N^k}$ toric code, where the class of allowed loop configurations, i.e. the local branching constraint, changes across the phase transition. Finally, we implement a phase transition between phases with the same topological order, but different fractionalization patterns for the anyonic excitations under a global symmetry~\cite{essin_classifying2013, chen_symmetry_2017, liu_simulating2024}. These are distinct phases as long as the symmetry is preserved, and is referred to as symmetry enriched topological order. We can tune from the $\mathbb{Z}_4$ toric code with trivial fractionalization pattern for all excitations under a global anti-unitary $\mathbb{Z}_2^T$ symmetry $\{ 1, \mathcal{K}X^{2} \} $, where $\mathcal{K}$ is complex conjugation, to another state where the vertex excitation $e_4$ exhibits non-trivial symmetry fractionalization (purple to turquoise). They are distinguished by a membrane order parameter $\langle O_{e_4} \rangle$ which is zero for non-trivial fractionalization~\cite{pollmann_detection_2012, zaletel_detecting_2014, huang_detection_2014}. 

The correlation length $\xi$ in the time-like direction $y$ diverges at each critical point, at which certain transition probabilities $p(cd\vert ab)$ of the stochastic circuit vanish and a higher symmetry is realized; see Fig.~\ref{fig:Path}. In particular, for the considered deformations we find that the stochastic circuit exhibits a conserved $U(1)$ symmetry~\cite{liu_simulating2024, boesl_quantum2025}. This allows for a hydrodynamic description~\cite{Chaikin_Lubensky_1995}: In 1+1D circuits, the local density $q_j$ of the associated charge exhibits Gaussian scaling, showing algebraic decay of equal-site correlations in time $t$. As this charge can be constructed from the automata values as $q_j = \sum_{x \in j} (n_x \; \text{mod} \;k)$ for some $k \leq N/2$, correlations of the associated power of $Z$ operators in $y$ direction exhibit critical scaling $\langle Z^k(x,y) (Z^\dagger)^k(x, y +r) \rangle \sim r^{-1/2}$ as well, leading to the observed diverging correlation length $\xi$. More generally, other correlators, such as loop-loop correlators, will also follow an algebraic decay in this direction, as we showed that they map to diagonal operators in the automaton. In SM~\cite{supp} Sec. F, we illustrate how conservation of higher multipole moments in the stochastic evolution inspire isoTNS with modified exponents, and how they lie at the endpoints of paths of critical states. Other classes of critical points exist as well. They do not have to possess emergent $U(1)$ conservation laws, but could also describe decoupled GHZ states~\cite{yu_dual2024}; in this case, only the fixed point of the automaton, not necessarily the update rule itself respects an additional symmetry.

From the parametrized isoTNS, we construct the parent Hamiltonian (see SM~\cite{supp}, Sec. C), which in general is not unique. We can construct continuous parent Hamiltonians for all paths between different phases as described by our approach. For critical points separating phases of the same topological order but with distinct symmetry fractionalization ($|\psi_{\text{crit},3}\rangle$ in the path we consider in Fig.~\ref{fig:Path}), this is already achieved by the imaginary time evolution of the fixed point Hamiltonians; for two distinct types of topological order ($|\psi_{\text{crit},1}\rangle$, $|\psi_{\text{crit},2}\rangle$), it involves a slightly more involved procedure as detailed in SM~\cite{supp} Sec. C. In all cases, the Hamiltonian stays topological and exhibits a non-trivial closing of the gap at the critical point.

%The parent Hamiltonian is in general discontinuous at critical points separating two distinct types of topological order ($|\psi_{\text{crit},1}\rangle$, $|\psi_{\text{crit},2}\rangle$ in the path we consider in Fig.~\ref{fig:Path}). Yet, they remain continuous for critical points separating phases of the same topological order but with distinct symmetry fractionalization ($|\psi_{\text{crit},3}\rangle$). From the perspective of state preparation, the properties of the parent Hamiltonian are, however, not necessarily important, as these states are not protected by a Hamiltonian but rather need to be actively error corrected. 

\begin{figure}
    \includegraphics[width=.98\columnwidth]{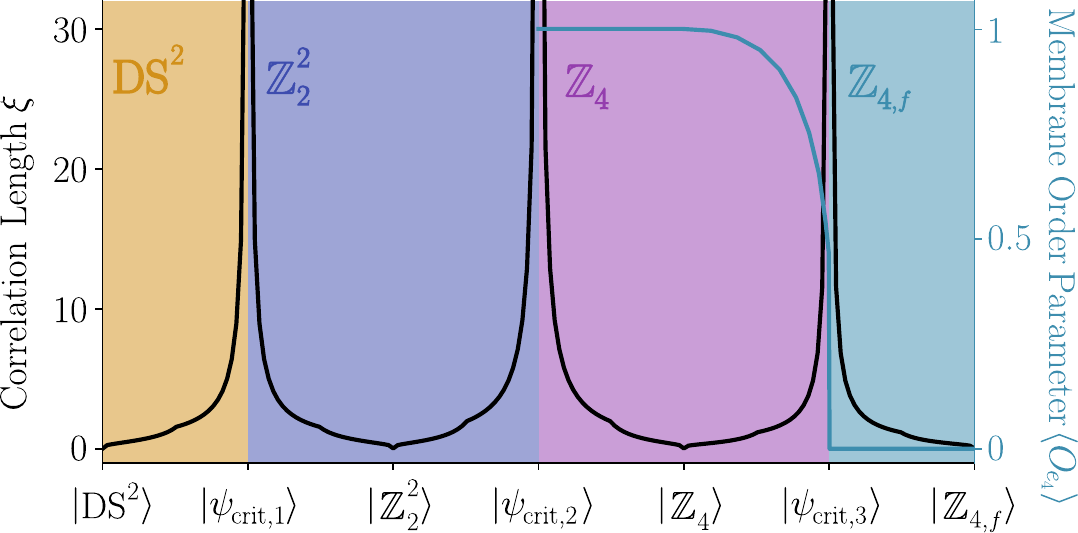}
    \caption{\label{fig:Path}
        \textbf{Path crossing multiple string-net phases.} We track the correlation length $\xi$ along the path shown in \fig{fig:Skeletons}; the tensor network passes from a pair of decoupled double-semion models (yellow) to a pair of $\mathbb{Z}_2$ toric codes (violet), which are subsequently coupled to a $\mathbb{Z}_4$ toric code (purple). If we further conserve a global anti-unitary $\mathbb{Z}_2^T$ symmetry, this phase is distinguished from another phase with the same topological order but non-trivial symmetry fractionalization on the anyons labeled $\mathbb{Z}_{4,f}$ (turquoise). The correlation length is zero at the fixed points and diverges at the critical points, where a conservation law emerges in the associated stochastic circuit, leading to algebraic decay of correlations in one direction. We also show a membrane order parameter $\langle O_{e_4} \rangle$ which distinguishes $\mathbb{Z}_4$ and $\mathbb{Z}_{4,f}$.
    }
\end{figure}

\textbf{\emph{Discussion and Outlook.---}} We have constructed skeletons of isometric tensor network states (isoTNS) that provide an organizing principle for abelian string-net phases and can be efficiently realized on quantum processors. They contain analytically tractable examples of continuous phase transitions beyond the anyon condensation framework. By mapping to classical automata we showed that polynomially many generalized Pauli strings (of arbitrary weight) can be efficiently measured classically for these isoTNS skeletons. 
The classical simulability is surprising as these states are generically neither generated by Clifford circuits nor match gates, which are familiar examples that admit efficient classical simulation. The analytical insights we derive for this class of complex quantum many-body states render them as intricate testbeds for quantum processors. 

Our framework suggests several generalizations, including modified variants of $\mathbb{Z}_N$ toric code states~\cite{watanabe_ground2023}. Working out how to connect non-abelian string-nets to stochastic processes will further enable the study of novel deformations of the fixed points as well as phase transitions between non-abelian topological orders~\cite{xu_characterization2021}; the known isometric tensor networks and sequential quantum circuits for the fixed points of general string-net models may serve as useful starting points~\cite{soejima_isometric2020, liu:2022}. Higher-dimensional stabilizer codes for topological and fracton order can be similarly treated~\cite{boesl_quantum2025}. Our class of states with guaranteed topological order can also be efficiently variationally optimized and may be useful in ground states searches for perturbed string-net Hamiltonians. We have also constructed continuous parent Hamiltonians in the vicinity of the phase transitions between different topological orders, which could inspire novel field theoretic constructions to describe topological criticality. 

Furthermore, the reverse mapping shows that any stochastic process with local rules defines an isoTNS, providing a systematic route to new, efficiently preparable quantum states with nontrivial correlations~\cite{haag_typical_2023, malz_computational_2024, boesl_2026}. Extending this approach to non-equilibrium stochastic processes, which violate detailed balance and can exhibit genuine non-equilibrium phase transitions, could uncover a richer class of novel critical phenomena~\cite{hinrichsen_nonequilibrium2000, odor_universality2004}. The equivalence discussed here is a special case of a more general relation between isoTNS and trace-preserving quantum operations; the virtual symmetries discussed in the context of topological order correspond to strong symmetries of the quantum channel, which enforce non-trivial steady-state degeneracy~\cite{ziereis_strong2025}. This perspective promises to shed more light on the remarkable stability of topological order in isoTNS~\cite{lu:2025}. 

\textit{\textbf{Acknowledgments.---}}We thank Wen-Tao Xu for insightful discussions. We acknowledge support from the Deutsche Forschungsgemeinschaft (DFG, German Research Foundation) under Germany’s Excellence Strategy–EXC–2111–390814868, TRR 360 – 492547816 and DFG grants No. KN1254/1-2, KN1254/2-1, the European Research Council (ERC) under the European Union’s Horizon 2020 research and innovation programme (grant agreement No 851161), the European Union (grant agreement No 101169765), as well as the Munich Quantum Valley, which is supported by the Bavarian state government with funds from the Hightech Agenda Bayern Plus. YJL acknowledges support from the MIT Center for Theoretical Physics - a Leinweber Institute, NSF CIQC Award No. 10434 and NSF Early Career DMR-2237244. 

\textit{\textbf{Data availability.---}}Data, data analysis, and simulation codes are available upon reasonable request on Zenodo~\cite{zenodo}.

\let\oldaddcontentsline\addcontentsline
\renewcommand{\addcontentsline}[3]{}
\bibliography{references}
\let\addcontentsline\oldaddcontentsline

\newpage
\leavevmode \newpage

\setcounter{equation}{0}
\setcounter{page}{1}
\setcounter{figure}{0}
\renewcommand{\thepage}{S\arabic{page}}  
\renewcommand{\thefigure}{S\arabic{figure}}
\renewcommand{\theequation}{S\arabic{equation}}
\onecolumngrid
\begin{center}
\textbf{Supplemental Material:}\\
\textbf{\papertitle}\\ \vspace{10pt}
\vspace{6pt}
Julian Boesl$^{1,2}$, Yu-Jie Liu$^{3,1,2}$, Frank Pollmann$^{1,2}$ and Michael Knap$^{1,2}$ \\ \vspace{6pt}
$^1$\textit{\small{Technical University of Munich, TUM School of Natural Sciences, Physics Department, 85748 Garching, Germany}} \\
$^2$\textit{\small{Munich Center for Quantum Science and Technology (MCQST), Schellingstr. 4, 80799 M{\"u}nchen, Germany}} \\
$^3$\textit{\small{Center for Theoretical Physics - a Leinweber Institute, Massachusetts Institute of Technology, Cambridge, MA 02139, USA}} \\
\vspace{10pt}
\end{center}

\maketitle

\tableofcontents

\section{A. Tensor Networks and Parent Hamiltonians for Abelian String-Net Models}
In this section, we discuss the precise tensor network construction of abelian string-net models, as well as possible parent Hamiltonians. For a rigorous derivation of abelian string-net models, we refer to Ref.~\cite{lin_generalizations2014}; here, we restrict ourselves to stating some general properties for reference.

String-net models are normally defined on trivalent graphs such as the honeycomb lattice; the associated wavefunction can be seen as a superposition of many different configuration of degrees of freedoms called strings. Each string-net model comes with a set of strings on the graph edges and branching rules, which determine which string combinations can meet on each vertex. Strings can be oriented, with each string type $a$ having a string of opposite direction $a^*$; a string is non-oriented if $a = a^*$. If the strings are labeled by elements of a finite abelian group $G$, a string $c$ on an edge is uniquely determined by the two other strings $a$ and $b$ around an adjacent vertex via the branching rule $a + b + c= 0$, where $+$ is the group operation, with all string being outwardly oriented. The string type opposite to $a$ is given by the group inverse, $a^* = -a$. Abelian string-nets are further defined by a set of complex functions $\{d_a, F(a,b,c), \gamma_a, \alpha(a,b)\}$, where $a,b,c$ are different types of strings; they define the amplitudes of different string configurations. $d_a$ is the quantum dimension of string $a$, whereas the $F$-symbol determines the phase change when ``rewiring" configurations locally; $\gamma_a$ and $\alpha(a,b)$ are likewise additional objects which determine relations between different string-net amplitudes~\cite{levin_string2005, soejima_isometric2020}. More abstractly, this data can be connected to unitary fusion categories~\cite{bonderson_interferometry2008}. Importantly, the functions $\{d_a, F(a,b,c), \gamma_a, \alpha(a,b)\}$ have to fulfill certain consistency relations to represent a valid abelian string-net; nevertheless, for a given topological order, the choice of these functions is not unique; different prescriptions lead to different fixed points in the same topological phase of matter.

Here, we focus on the case $G = \mathbb{Z}_N$; the arguments offer a straightforward generalization to products of cyclic groups $G = \mathbb{Z}_{N_1} \times \cdots \times \mathbb{Z}_{N_k}$, where the branching rule must be fulfilled for each cyclic group individually. As every finite abelian group is isomorphic to such a direct product~\cite{lin_generalizations2014}, this approach covers all abelian string-net models.

The unique branching rule of abelian string-nets implies that one edge direction on the honeycomb is redundant, as the information about the string on these edges is already contained in the string configurations on the edges in the other two directions. These models can thus be mapped to a square lattice without any issues. We will therefore describe the tensor networks for string-net states in terms of the square lattice; the original honeycomb can easily be restored by adding the third physical degree of freedom. To define tensor networks for abelian $\mathbb{Z}_N$ string-net states, we put an $N$-dimensional qudit on every edge of a square lattice, where the local degrees of freedom are labeled by $0, \dots, N-1$. The different physical states can be seen as different types of strings. The abelian string-net state is an equal-weight superposition of all string configurations which fulfill the local branching constraint
\begin{equation}
    \includegraphics[scale=0.425]{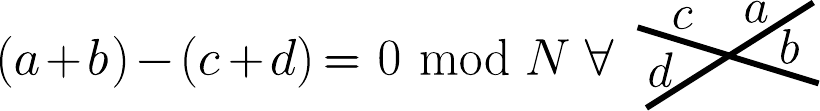}
    \label{eq:Branching}
\end{equation}
on every vertex $v$. This branching rule on the square lattice arises from the honeycomb rule by considering two nearest-neighbor honeycomb vertices and ``contracting" the edge connecting them, leading to a square lattice geometry; the qudits on edges $a$ and $b$ describe outwardly oriented strings relative to $v$, while the qudits on $c$ and $d$ describe inwardly oriented strings. Then, the string-net wavefunction $\ket \psi$ is of the form $\ket\psi \propto \sum_c \alpha_c \ket c$ with $\vert \alpha_c \vert = 1$, where $\ket c$ is a string-net configuration (see \eq{eq:PsiDef} for a graphical depiction). We set the amplitude of the vacuum configuration $\ket {\text{vac}} = \ket {0 \cdots 0}$ to one, $\alpha_0 = 1$. In general, other amplitudes will be complex $\alpha_c \in \mathbb{C}$; the phase factors of the configurations depend on the topological order and in particular on the underlying data $\{d_a, F(a,b,c), \gamma_a, \alpha(a,b)\}$. For every pair of $\mathbb{Z}_N$ fixed points $\ket \psi$ and $\ket {\psi^\prime}$, a global diagonal unitary $U$ exists which transforms one fixed point into the other $\ket {\psi^\prime} =U\ket {\psi} $; its action merely consists of replacing the complex phases of the amplitudes $\alpha_c$ for all configurations $\ket c$, $U\ket c = \alpha^\prime_c\alpha_c^* \ket c$.

As in the main text, we introduce a tensor $T_{abcd}^{\sigma\rho}$ at every vertex, with virtual legs $a,b,c,d$ of bond dimension $D$ and physical legs $\sigma,\rho$ on two edges. The tensor network state is obtained by contracting all virtual legs as
\begin{equation}
        \ket{\psi} = \sum_{\sigma_1, \cdots, \sigma_k} \text{tTr}\left( \left\{ T^{\sigma_1\sigma_2}, \cdots, T^{\sigma_{k-1}\sigma_k} \right\} \right) \ket{\sigma_1\cdots\sigma_k},
    \label{eq:TensorWFSM}
\end{equation}
where $\text{tTr}$ denotes this tensor contraction. We discuss two classes of tensor networks to describe abelian string-net models: \emph{Single-} and \emph{double-line} TNS. \emph{Single-line} TNS couple physical and virtual legs and describe states where the phases $\alpha_c$ depend on local properties, as is the case for the $\mathbb{Z}_N$ toric code~\cite{chen_tensor2010}. A single-line tensor is decomposed as 
$T_{abcd}^{\sigma\rho} = \sum_{a^\prime b^\prime} \delta_{aa^\prime}^\sigma \delta_{bb^\prime}^\rho W_{(a^\prime b^\prime )(cd)}$, or graphically
\begin{equation}
    \includegraphics[scale=0.65]{SingleLine.pdf},
    \label{eq:SingleLineSM} 
\end{equation}
where $W$ is an $N^2 \times N^2$ matrix and $\delta$ is a ``plumbing" tensor with $\delta_{ab}^\sigma = 1$ if $a = b = \sigma$ and zero otherwise. The bond dimension is therefore equivalent to the physical dimension $D = N$. In particular, this state represents the $\mathbb{Z}_N$ toric code for
\begin{equation}
    \vert W_{(ab)(cd)}\vert^2 = \frac{1}{N} \quad \text{if} \; (a+b) -(c+d) = 0 \; \text{mod} \; N
    \label{eq:WSingleLine}
\end{equation}
and zero otherwise. This prescription simply enforces the branching constraint from \eq{eq:Branching} and the equal-weight condition for all allowed configurations. If all non-zero entries are real and positive, this state is the ground state of the standard $\mathbb{Z}_N$ toric code Hamiltonian. Other valid choices correspond to wavefunctions which are connected by a finite-depth diagonal unitary circuit. They are thus in the same topological phase, while they might still be distinguished e.g.~by fractionalization patterns in the presence of additional symmetries.

A more general set which describes all abelian string-net models including the aforementioned ones are \emph{double-line} TNS, whose structure represents how string-nets are built from fusions of plaquette loops~\cite{gu_tensor2008, gu_tensor2009, buerschaper_explicit2009, soejima_isometric2020}.
Double-line states allow for non-local phase factors as present in ``non-toric code" string-net models~\cite{levin_braiding2012, zhang_nonhermitian2020, xu_tensor2018}. The tensor has bond dimension $D = N^2$ and is decomposed into $T_{(ab)(cd)(a^\prime b^\prime)(c^\prime d^\prime)}^{\sigma\rho} = \sum_{kqhl} \delta_{ak,bq}^\sigma \delta_{ch,dl}^\rho W_{(kq,hl)(a^\prime b^\prime,c^\prime d^\prime)}$, or graphically 
\begin{equation}
    \includegraphics[scale=0.55]{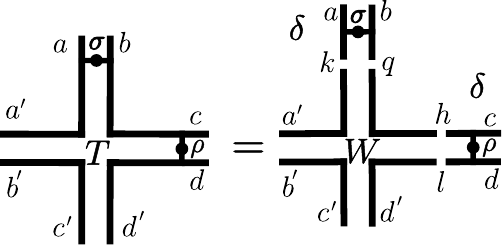},
    \label{eq:DoubleLine} 
\end{equation}
where the $\delta$-tensor has a domain-wall structure, $\delta_{ak,bq}^\sigma = 1$ if $a = k$, $b = q$ and $\sigma = (a -b) \; \text{mod} \; N$. $W$ is a $N^4 \times N^4$ matrix with the form
\begin{equation}
    W_{(ab,cd)(a^\prime b^\prime,c^\prime d^\prime)} = A_{acd^\prime b^\prime} \delta_{aa^\prime}\delta_{cb}\delta_{c^\prime b^\prime}\delta_{dd^\prime},
    \label{eq:WDLConstraint}
\end{equation}
where the tensor $A$ encodes the non-zero elements. This enforces the closed-loop character. The strings of the adjacent plaquette loops are fused to obtain the physical degree of freedom. At the fixed points, the entries have the same magnitude
\begin{equation}
    \vert A_{acd^\prime b^\prime} \vert^2 = \frac{1}{N} \quad \text{for all} \; a,c, b^\prime, d^\prime.
    \label{eq:ADoubleLine}
\end{equation}
The complex phases of the entries are determined by the topological order through the string-net data $\{d_a, F(a,b,c), \gamma_a, \alpha(a,b)\}$. If they are all real and positive, the tensor network again represents the standard $\mathbb{Z}_N$ toric code; every other $\mathbb{Z}_N$ string-net model can also be represented in this form. Double-line tensors are in fact a subclass of more general triple-line tensors which can represent any string-net model~\cite{shukla_boson2018, soejima_isometric2020}; for abelian string-net models, the third line is not required as the outcome of fusing two strings is unique, allowing for this simpler description.

\section{B. Virtual symmetries and deformations of abelian string-nets}
Topological order is stable against local perturbations, i.e., the ground state of the sum of a fixed point Hamiltonian and a weak local perturbation is in general still topologically ordered~\cite{kitaev_faulttolerant2003, klich_stability2010}. Actual tensor network deformations however are much more delicate: Given some fixed point $T_{abcd}^\sigma$ which represents a topologically ordered phase, a generic deformation $T_{abcd}^\sigma + \varepsilon (T^\prime)_{abcd}^\sigma$ will destroy topological order at any strength $\varepsilon \neq 0$. This is because such a deformation will act also on the virtual level of the tensor, thus corresponding to a non-local physical perturbation which implies condensation of some anyons~\cite{shukla_boson2018}. While such a state still has a local parent Hamiltonian, it cannot be connected smoothly to the fixed point Hamiltonian as $\varepsilon \rightarrow 0$~\cite{chen_tensor2010}.

However, it is possible to identify deformations which leave topological order intact when small enough, provided we keep the decomposition into a $W$-matrix and plumbing tensors. These have to preserve certain virtual symmetries of the fixed point tensors which protect the topological order~\cite{schuch_peps2010, shukla_boson2018, sahinoglu_characterizing2021}. In this section, we discuss the relevant symmetries for the different tensor network representations of abelian string-net models discussed in the main text. To do so, we introduce the clock operator $Z$ with $Z\ket{j} = e^{2\pi ij/N} \ket{j}$ and the shift operator $X$ with $X\ket{j} = \ket{(j + 1 )\;\text{mod} \; N}$ as in the main text.

In the single-line case, the virtual symmetry in question directly represents the closed-loop constraint which can be expressed in terms of $Z$ operators acting on the virtual legs

\begin{equation}
    \includegraphics[scale=0.575]{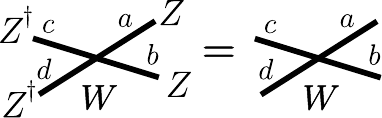},    \label{eq:ZVirtSymm}
\end{equation}
If this symmetry is broken by the modified tensor, open endpoints of some strings are allowed. Importantly, for a potentially small weight $\propto \varepsilon$ of a parity-violating configuration around a vertex, open strings at all lengths contribute with a weight of $\propto \varepsilon^2$. As these open strings are not suppressed in length, they induce condensation of the vertex excitations, which correspond to violations of the vertex stabilizers $\mathcal{A}_v$, and thus to a loss of topological order. 

The double-line fixed points tensors exhibit more symmetries. The first class is given by the delta functions in \eq{eq:WDLConstraint}. They can equally be expressed in terms of $Z$ operators on the relevant legs and ensure the absence of open loops. However, they may be violated without destroying the topological order: In contrast to the single-line case, they can only induce short-length open strings. Longer strings necessitate a number of violations which scales with string length $L$ and are thus suppressed exponentially $\propto \varepsilon^L$. This guarantees that the topological order remains stable for small deformations~\cite{shukla_boson2018}. Nevertheless, in this work we only consider deformations which conserve these symmetries. The reason for this lies with the other important ingredient of our mapping to stochastic circuits: The isometry condition for double-line tensors tacitly assumes that the closed loop constraints of \eq{eq:WDLConstraint} are always conserved; otherwise, the tensor network cannot be efficiently contracted~\cite{liu_simulating2024}. Accordingly, if we allow these additional configurations in the double-line tensor, there will in general be no mapping to an equivalent single-line tensor anymore as these cannot support non-condensed vertex excitations as discussed above. This also precludes a mapping to a stochastic automaton.

The symmetry which actually protects the topological order is a loop of $X$ operators, potentially accompanied by some diagonal operator, graphically expressed as 

\begin{equation}
    \includegraphics[scale=0.575]{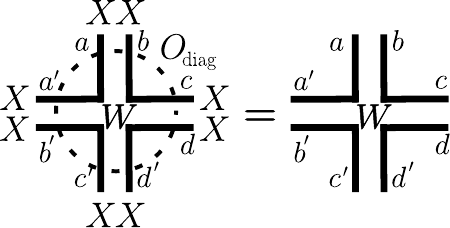}. 
    \label{eq:XVirtSymm}
\end{equation}
This diagonal operator $O_{\text{diag}}$ depends on the fixed point and the topological order in question. It at most adds a complex phase, so $\|O_{\text{diag}}\| = 1$. For the regular toric code fixed point with only positive weights it is trivial, $O_{\text{diag}} = 1$. A deformation which does not respect this symmetry brings condensation of the plaquette excitations with it, as here strings of $Z$ operators at all lengths contribute equally. On the level of the tensor, this condition enforces that entries which correspond to the same physical configuration around a vertex have the same magnitude.

The difference between those symmetries which may be violated and those which have to be respected to protect the topological order can be formalized by splitting the vector space of possible deformations of the fixed point tensor into different subspaces~\cite{shukla_boson2018}. Symmetries which are allowed to be violated define deformations which collapse the tensor network upon replacing one local tensor; they thus need to appear in adjacent pairs. A pair of excitations at far distances thus needs a number of deformed tensors scaling with said distance, precluding their proliferation. These deformations are said to lie outside the stand-alone subspace of the fixed point tensor, which consists of all directions which are stable under insertion of a single tensor. For the single-line tensor, this is the entire vector space of possible directions; for the double-line tensors it is given by those which conserve equivalence of two legs meeting at a corner such as $a$ and $a^\prime$. The deformations inside this subspace which do not destroy topological order at arbitrary strength lie inside another subspace called the MPO-injective subspace. This space is protected by virtual matrix product operators, i.e. the loops of $Z$ operators for single-line tensors and the loops of $X$ operators for double-line tensors. 

In the MPO-injective subspace, all deformations can be mapped to local physical perturbations~\cite{chen_tensor2010, sahinoglu_characterizing2021, shukla_boson2018}. This is the space our paths lie in: They change the weights of local string-net configurations so that in general $\vert \alpha_c\vert \neq 1$; however, no new configurations with open strings are introduced. The physical deformations associated to these deformations are thus diagonal in the $Z$ basis. The deformed state $\ket{\psi}$ and the fixed point $\ket{\psi_{\text{fix}}}$ are connected by an imaginary time evolution $\mathcal{T}$
\begin{equation}
    \ket{\psi} =\mathcal{T} \ket{\psi_{\text{fix}}} = \prod_v e^{\sum_j \beta_j P_v^j}\ket{\psi_{\text{fix}}},
    \label{eq:ImaginaryTime}
\end{equation}
where $P_v^j$ is a projector on some diagonal configuration at vertex $v$ and the coefficients $\beta_j$ are determined by the deformation. While the fixed point state $\ket{\psi_{\text{fix}}}$ has zero correlation length $\xi = 0$ due to the commuting projector nature of its parent Hamiltonian, the perturbation induces a non-vanishing correlation length $\xi \neq 0$.

\section{C. Parent Hamiltonians for deformed string-nets}
Each fixed point of an abelian string-net model is the ground state of a Hamiltonian on a square lattice of the form
\begin{equation}
    H = \sum_v \mathcal{A}_v + \sum_p \mathcal{B}_p \mathcal{P}_p,
    \label{eq:FPHamiltonian}
\end{equation}
where the sums run over the vertices $v$ and the plaquettes $p$ of the lattice. The vertex terms $\mathcal{A}_v$ enforce the branching rules by penalizing states which do not fulfill the fusion constraint. Introducing the operator $A_v$ as

\begin{equation}
    \includegraphics[scale=0.425]{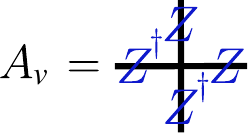},
    \label{eq:ADef}
\end{equation}
we can write the vertex operator as $\mathcal{A}_v = \left (1- \frac{1}{N} \sum_{k=0}^{N-1} (A_v)^k \right)$. From $A_v \ket {\psi_{\text{fix}}} = \ket{\psi_{\text{fix}}}$, this definition ensures that $\mathcal{A}_v$ is a projector with $\mathcal{A}_v \ket {\psi_{\text{fix}}} = 0$ if the branching constraint is fulfilled at vertex $v$.

The plaquette terms $\mathcal{B}_p$ induce dynamics between different configurations. Similar to the vertex term, we introduce an operator $B_p$. This operator is supported on the 12 degrees of freedoms living on the edges around the vertices of the plaquette~\cite{levin_string2005, lin_generalizations2014}. It can be decomposed into a diagonal part $O_{\text{ph}}$, which acts on all qudits, and an off-diagonal part $(XX^\dagger)^{\otimes2}$ on the edges of the plaquette as

\begin{equation}
    \includegraphics[scale=0.425]{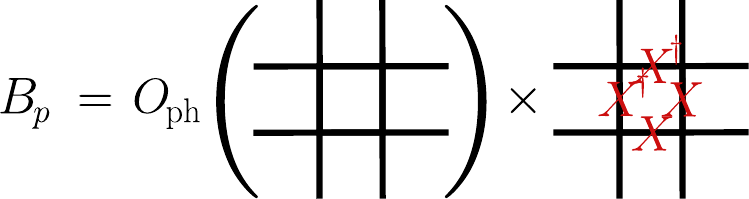}.
    \label{eq:BDef}
\end{equation}
The operator $O_{\text{ph}}$ is determined by the complex functions $\{d_a, F(a,b,c), \gamma_a, \alpha(a,b)\}$ and simply allocates a phase to a configuration, so $\|O_{\text{ph}}\| = 1$. The plaquette term is again a superposition of the powers as $\mathcal{B}_p = \left ( 1- \frac{1}{N} \sum_{k=0}^{N-1} (B_p)^k \right)$. The projector $\mathcal{P}_p = \prod_{v\in p} (1-\mathcal{A}_v)$, where $v\in p$ indicates that the vertex $v$ is adjacent to plaquette $p$, restricts its action to the closed-loop subspace. As $O_{\text{ph}}$ is chosen so that $B_p \ket {\psi_{\text{fix}}} = \ket {\psi_{\text{fix}}}$ on the string-net fixed point $\ket {\psi_{\text{fix}}}$, this combined operator is also a projector with $\mathcal{B}_p \mathcal{P}_p\ket {\psi_{\text{fix}}} = 0$.

Moreover, the operators $\mathcal{A}_v$ and $\mathcal{B}_p\mathcal{P}_p$ form a set of mutually commuting operators. The Hamiltonian is thus exactly solvable and has a ground state which minimizes each projector; this state can be written as $\ket{\psi_{\text{fix}}} \propto \prod_p \left( \sum_{k=0}^{N-1} (B_p)^k \right)\ket{0\cdots0}$. The construction of this ground state can be seen as subsequently applying the smallest possible closed loops on each plaquette and fusing them according to the rules given by $\{d_a, F(a,b,c), \gamma_a, \alpha(a,b)\}$~\cite{gu_tensor2009, buerschaper_explicit2009}.

Excitations correspond to violations of the local projectors, while the ground state as defined above represents their vacuum. They can be created only in pairs and represent a set of $N^2$ abelian anyons with fusion rules which can be derived from $\{d_a, F(a,b,c), \gamma_a, \alpha(a,b)\}$. Accordingly, $A_v$ represents a minimal braiding process of a plaquette anyon around a vertex, while $B_p$ represents a minimal braiding process of a vertex anyon around a plaquette.

From the imaginary time evolution $\mathcal{T}$ as defined in \eq{eq:ImaginaryTime} and the fixed point Hamiltonian, we can also construct one possible parent Hamiltonian for the deformed state $\ket \psi = \mathcal{T} \ket{\psi_{\text{fix}}}$. As no open string configurations are introduced, the deformed state is still annihilated by every vertex operator $\mathcal{A}_v$. Introducing $\tilde{B}_p = \mathcal{T}B_p\mathcal{T}^{-1}$, it is easily seen that the deformed state $\ket \psi$ is an eigenstate of $\tilde{\mathcal{B}}_p = \left( 1- \frac{1}{N} \sum_{k=0}^{N-1} (\tilde{B}_p)^k \right)$ with zero eigenvalue $\tilde{\mathcal{B}}_p \ket \psi = 0$. Because $\mathcal{T}$ is a product of local diagonal operators, $\tilde{\mathcal{B}}_p$ has the same support as the original plaquette term $\mathcal{B}_p$. We can thus write down a frustration-free, local parent Hamiltonian of $\ket \psi$ as

\begin{equation}
    H = \sum_v \mathcal{A}_v + \sum_b (\tilde{\mathcal{B}}^\dagger_p\tilde{\mathcal{B}}_p )\mathcal{P}_p.
    \label{eq:DeformedHamiltonian}
\end{equation}
The plaquette terms $(\tilde{\mathcal{B}}^\dagger_p\tilde{\mathcal{B}}_p )\mathcal{P}_p$ of the deformed Hamiltonian are positive semi-definite; however, in general they are not projection operators anymore. By a suitable normalization, which amounts to an additional local operator in the term, the projector structure can be restored. Furthermore, the plaquette operators do not commute between each other anymore; this induces a non-vanishing correlation length and spoils the exact solvability of the deformed Hamiltonian, as we cannot express the entire spectrum in simple terms anymore.

For a path connecting two fixed points of different orders, an important question is whether we can find a parametrized parent Hamiltonian which remains continuous along the entire path. This question is in fact non-trivial: The parent Hamiltonians of perturbed tensors $T + \varepsilon T^\prime$ which break the virtual symmetries of the fixed point $T$ cannot be connected continuously to its topological Hamiltonian as $\varepsilon \rightarrow 0$; rather, the Hamiltonian at $\varepsilon = 0$ is trivial and goes under the name of ``uncle Hamiltonian"~\cite{cirac_matrix2021, chen_tensor2010}. By contrast, for a path of isoTNS skeletons, we show in the following how to construct a continuous parent Hamiltonian interpolating between the two topological Hamiltonians.

For transitions between phases which differ only in their symmetry fractionalization pattern, such a parametrization is already achieved by the prescription given in \eq{eq:DeformedHamiltonian}, as has been discussed in~\cite{liu_simulating2024}. The deformed Hamiltonians coming from the two fixed points coincide at the critical point, thus the Hamiltonian obtained from an imaginary time evolution from the respective fixed point can be used as a continuous interpolation.

For critical points between phases with different topological order, the situation is more complicated: Here, the two parent Hamiltonians as given by \eq{eq:DeformedHamiltonian} will differ in general at the critical point, leading to a discontinuity. However, a continuous parent Hamiltonian path does exist, and can be constructed from the Hamiltonian (\ref{eq:DeformedHamiltonian}). We first consider two topological phases with the same local branching constraint, for instance between the toric code and the double semion model; in this case, the vertex projectors $\mathcal{A}_v$ remain unchanged in the deformation, thus the only discontinuity comes from the deformed plaquette operators $(\mathcal{B}^\dagger_p \mathcal{B}_p) \mathcal{P}_p$. As this operator acts non-diagonally only on the four qudits directly adjacent to the plaquette, we can split it up with respect to the configuration of the eight outer qudits which remain unchanged, $(\mathcal{B}^\dagger_p \mathcal{B}_p) \mathcal{P}_p = \sum_\text{env} \Bigl((\mathcal{B}^\dagger_p \mathcal{B}_p) \mathcal{P}_p \mathcal{P}_\text{env} \Bigr)$, where the sum runs over all $N^8$ environment configurations. Accordingly, each operator $\Bigl((\mathcal{B}^\dagger_p \mathcal{B}_p) \mathcal{P}_p \mathcal{P}_\text{env} \Bigr)$ couples $N$ different local states. For certain choices of $\text{env}$, this $N$-state operator is the same from both sides, therefore it does not need any further treatment. For other configurations, it differs; this is the case if the imaginary time evolution reduces the weight of all $N$ configurations equally, leaving the operator $\Bigl((\mathcal{B}^\dagger_p \mathcal{B}_p) \mathcal{P}_p \mathcal{P}_\text{env} \Bigr)$ unchanged along the entire path to the critical point. To amend this, we reduce the weight of these projectors in the Hamiltonian when approaching the critical point, i.e. for a path parametrized by $g \in [-1,1]$ with $H(g = -1)$ and $H(g = 1)$ the fixed point Hamiltonians, while $H(g = 0)$ is the critical Hamiltonian, we replace $\Bigl((\mathcal{B}^\dagger_p \mathcal{B}_p) \mathcal{P}_p \mathcal{P}_\text{env} \Bigr) \rightarrow \vert g \vert \Bigl((\mathcal{B}^\dagger_p \mathcal{B}_p) \mathcal{P}_p \mathcal{P}_\text{env} \Bigr)$. Using this prescription, the modified Hamiltonian remains continuous along the entire path.

\begin{figure*}
    \includegraphics[width=0.4\textwidth]{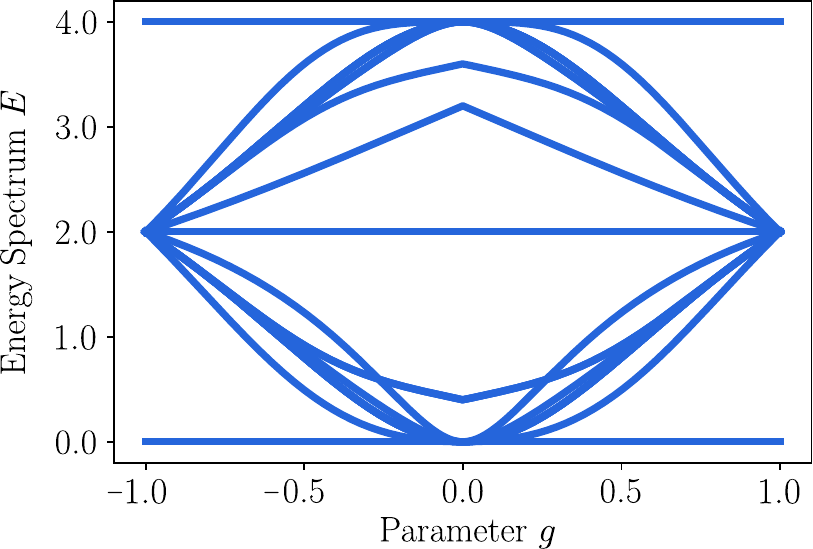}
    \caption{\label{fig:TCDSSpectrum}
        \textbf{Spectrum of a Parent Hamiltonian.} Energy spectrum of the continuous parent Hamiltonian on the isoTNS path (\ref{eq:PathTCDS}) between the double semion fixed point $\ket{\text{DS}}$ at $g = -1$ and the toric code fixed point $\ket{\mathbb{Z}_2}$ at $g = 1$. The spectrum is obtained from exact diagonalization on a periodic lattice with $2 \times 2$ plaquettes and 12 qubits. Away from the critical point $g = 0$, the ground state manifold is fourfold degenerate and gapped, while at this critical point the lowest lying excitations became gapless.
    }
\end{figure*}

This approach can be generalized to transitions between string-nets with different branching rules, i.e the $\mathbb{Z}_2^2 - \mathbb{Z}^4$ transition in \fig{fig:Path}. The plaquette operators can be treated in the same way,  by reducing the weight of all parts which lead to the discontinuity at the critical point. In particular, this is the case for all environments where the $N$ possible 12-qudit configurations all violate the branching rules of the other fixed point. At the same time, here the vertex projectors $\mathcal{A}_v$ also differ between the two fixed points and need to be modified. This is done in the following way: At each fixed point, the operator $\mathcal{A}_v$ annihilates configurations respecting the associated branching constraint ($\mathcal{A}_v \ket\psi = 0$), while it penalizes those violating it ($\mathcal{A}_v \ket \psi = \ket \psi)$. In our modified parent Hamiltonian, we demand that the energy penalty of a configuration which violates the branching constraint of the fixed point of the current phase, but which respects the different constraint of the other fixed point, be reduced on approaching the critical point, i.e for such a configuration $\mathcal{A}_v (g) \ket \psi = \vert g \vert\ket \psi$. Thus, at the critical point, the vertex projectors only penalize those configurations which violate both branching constraints.

From this, a continuous Hamiltonian for any path connecting two abelian string-net fixed points can be constructed. Importantly, the ground state degeneracy remains unchanged away from the critical point, and the Hamiltonian remains non-trivial at the critical point, i.e. a gapless continuum above the ground state manifold emerges, which we have checked explicitly numerically for transitions such as the TC-DS via exact diagonalization of small systems; in \fig{fig:TCDSSpectrum}, we show the energy spectrum of the continuous parent Hamiltonian on this path for a system of $2 \times 2$ plaquettes with 12 qubits.

\section{D. Mapping for off-diagonal operators}
We argue in the main text that all generalized Pauli strings can be accessed via sampling of a classical automaton for the isoTNS we describe. For example, for $N=2$ those are Pauli strings of any weight, containing $Z, X$ and $ZX = i Y$ operators; for larger $N$ the generalized Pauli strings are strings of $Z^a X^b, \forall a,b \in \{0,N-1\}$ and $a + b > 0$ (i.e., excluding the identity in the generalized Pauli string, so that the number of operators sets its weight). To this end, we associate to any generalized Pauli string $O$ a diagonal operator $O_{\text{diag}}$ such that $\bra\psi O \ket\psi = \bra\psi O_{\text{diag}} \ket\psi$, where $\ket \psi$ is either (i) a generic single-line tensor network state or (ii) a double-line tensor network state which respects the virtual symmetries of an abelian string-net state. In particular, this operator $O_{\text{diag}}$ can be decomposed into a product of local diagonal operators of fixed support, allowing for efficient evaluation in our stochastic circuit mapping if the $W$-matrix is additionally normalized and the state thus an isoTNS.
We work out this mapping exactly for a product of $X$ operators in case (i). Furthermore, we will discuss how the closed-loop constraint of abelian string-nets only allows certain operators with non-vanishing expectation value, covering case (ii). Finding the diagonal operator corresponding to an arbitrary generalized Pauli string is straightforward as those are composed of products of $X$ and $Z$, the latter of which is diagonal anyways. We emphasize that even though all generalized Pauli strings can be efficiently measured, a general operator cannot. As an example consider the measurement of an arbitrary rotation on every site supporting the operator. Generally such an operator is composed of a number of generalized Pauli strings that scales exponentially with the support of the operator. Hence, classically exponentially many diagonal operators have to be evaluated, which is not efficient. Arbitrary local operators with sufficiently small support $k$, however, can be measured as they are composed of $N^{2k}$ contributions. To summarize, the following operators can be efficiently measured by mapping to classical stochastic automata: 
\begin{itemize}
    \item Generalized $k$-body Pauli strings $O = \prod_i O_i$ with $k$ arbitrary, and $O_i = Z_i^a X_i^b, \forall a,b \in \{0,N-1\}$ and $a + b > 0$.
    \item General operators with sufficiently small $k$-body support, by decomposing them into $N^{2k}$ generalized Pauli strings.
\end{itemize}

For single-line tensors with exclusively positive entries, the fact that a local operator expectation value can be mapped to a classical observable with possibly extended support has already been pointed out in Ref.~\cite{verstraete_criticality2006}. Here, we go beyond this observation by covering not only general single-line states, but also double-line states of abelian string-net models, as well as considering Pauli strings of arbitrary weight and how they can be mapped to a product of local diagonal operators.

We first consider a state $\ket \psi$ with $N$-level qudits living on the edges which is determined as a single-line tensor network with some $W$-matrix as given by \eqs{eq:TensorWFSM}{eq:SingleLineSM}. We want to evaluate the expectation value of $k$ $X$ operators applied on some edge degrees of freedom $\bra \psi \prod_{i=0}^{k-1} X_i  \ket \psi $. The plumbing tensor $\delta_{ab}^\sigma$ allows us to pull the $X$ operators from the physical legs of $\ket \psi$ to the virtual legs, as
\begin{equation}
    \includegraphics[scale=0.425]{SkeletonsPlumbing.pdf}.
    \label{eq:Plumbing}
\end{equation}
The state $\ket {\psi^\prime} = \prod_{i=0}^{k-1} X_i  \ket \psi$ is thus again a single-line tensor where at most $2k$ local $W$-matrices have been changed by application of the plumbed $X^\dagger$ operators. If the matrix $W_v$ at vertex $v$ is affected by at least one $X^\dagger$ operator, we designate the changed matrix as $W^\prime_v$. As $X^\dagger$ is a shift operator, we can write out this matrix entry-wise as $(W^\prime_v)_{(a^\prime b^\prime)(c^\prime d^\prime)} = (W_v)_{(a b)(c d)} $ where $a^\prime = (a-1) \; \text{mod} \; N$ if an $X^\dagger$ operator is applied to leg $a$ and $a^\prime = a$ otherwise, equally for the other legs.

The expectation value is thus equivalent to the overlap $\bra \psi {\psi^\prime}\rangle$ of two single-line tensor network states. The equivalence between physical and virtual degrees of freedom in single-line tensors implies that such a state can be written as a superposition of orthogonal states $c$ in the computational basis $\ket \psi = \sum_c \alpha_c \ket c$, where the amplitude $\alpha_c = \prod_{v, abcd \in c} (W_v)_{(ab)(cd)}$ is the product of the local $W$-matrix entries. The product of $X$ operators simply maps each configuration $c$ to another one $c^\prime$ as $\prod_i X_i \ket c = \ket {F(c)}$. The overlap is thus given by $\bra \psi {\psi^\prime}\rangle = \sum_c \alpha_{F(c)}^*\alpha_c$.

For any single-line tensor network state, the diagonal operator in question is therefore determined by the relation $O_{\text{diag}} \ket c = \frac{\alpha_{F(c)}^*}{\alpha_c^*} \ket c$ if $\alpha_c \neq 0$, as then

\begin{equation}
    \bra \psi O_{\text{diag}} \ket \psi = \sum_c \vert\alpha_c\vert^2 \bra c O_{\text{diag}} \ket c = \sum_c \alpha_{F(c)}^*\alpha_c = \bra \psi {\psi^\prime}\rangle = \bra \psi \prod_{i=0}^{k-1} X_i  \ket \psi.
\end{equation}
We split this operator into diagonal operators $O_v$ acting only on the four degrees of freedom around one affected vertex $v$. They are explicitly expressed in terms of the $W$-matrix as 

\begin{equation}
    O_v = \sum_{abcd, W_{(ab)(cd)} \neq 0} \left(\frac{W^\prime_{(ab)(cd)}}{W_{(ab)(cd)}} \right)^*\ket{\text{abcd}}\bra{\text{abcd}} + \sum_{abcd, W_{(ab)(cd)} = 0} \ket{\text{abcd}}\bra{\text{abcd}},
    \label{eq:DiagVertexOperator}
\end{equation}
where $\ket{\text{abcd}}$ is a state in computational basis of the local $N^4$-dimensional Hilbert space. From this, $O_{\text{diag}} = \prod_v O_v$ can be explicitly constructed.

The simple picture of this strategy is that instead of evaluating an off-diagonal operator which permutes configurations into each other $\ket c \rightarrow \ket {F(c)}$, we may as well consider a diagonal operator which permutes amplitudes $\alpha_c \rightarrow \alpha_{F(c)}$. Starting from the simple case of a product of $X$ operators, it becomes clear how to treat any generalized Pauli string, as those are decomposed of products of $X$ and $Z$ operators, and $Z$ operators are already diagonal in the configurational basis. If the $W$-matrix is normalized as given in \eq{eq:WNorm}, the operators $O_v$ and thus also $O_{\text{diag}}$ can be efficiently accessed by classical sampling.

The arguments in the preceding paragraphs hold for arbitrary single-line states and do not require any virtual symmetries as prerequisites. For the double-line states on the other hand, we will make use of the closed loop-constraint which we impose on all deformed string-net liquids; this dependence on the virtual symmetries is to be expected as the double-line states are more directly inspired by the underlying structure of string-net liquids. We make use of the fact that this structure of abelian string-net liquids strongly restricts which operators can have non-zero expectation value, as the expectation value of an operator which creates only open strings vanishes. Furthermore, from the discussion in Section C, we know that a string-net fixed point $\ket{\psi_{\text{fix}}}$ is an eigenstate of the simple plaquette operator $B_p = O_{\text{ph}} (XX^\dagger)^{\otimes 2}$ as $B_p \ket{\psi_{\text{fix}}} = \ket{\psi_{\text{fix}}}$, where $O_{\text{ph}}$ depends on the topological order. It is thus clear that for the smallest loop of $X$ operators around a plaquette the expectation value is given by $\bra{\psi_{\text{fix}}}(XX^\dagger)^{\otimes 2}\ket{\psi_{\text{fix}}} = \bra{\psi_{\text{fix}}}O_{\text{ph}}^{-1}\ket{\psi_{\text{fix}}}$. Moreover, all deformations $\ket \psi = \mathcal{T} \ket{\psi_{\text{fix}}}$ under an imaginary time evolution $\mathcal{T}$ that we consider also conserve the closed-loop condition. We can thus express the expectation value of an $X$ operator loop in terms of a diagonal operator as well by identifying

\begin{align}
    \bra{\psi}(XX^\dagger)^{\otimes 2}\ket{\psi} & = \bra{\psi}(XX^\dagger)^{\otimes 2}\mathcal{T} \ket{\psi_{\text{fix}}} = \bra{\psi}\left((XX^\dagger)^{\otimes 2}\mathcal{T} (X^\dagger X)^{\otimes 2} \right) (XX^\dagger)^{\otimes 2} \ket{\psi_{\text{fix}}} \nonumber \\ & = \bra{\psi}\left((XX^\dagger)^{\otimes 2} \mathcal{T}(X^\dagger X)^{\otimes 2} \right) O_{\text{ph}}^{-1}\ket{\psi_{\text{fix}}} \nonumber \\ & =\bra{\psi}\left((XX^\dagger)^{\otimes 2} \mathcal{T}(X^\dagger X)^{\otimes 2} \right) O_{\text{ph}} ^{-1} \mathcal{T}^{-1}\ket{\psi}.
    \label{eq:LoopMappingGeneral}
\end{align}
The three operators $\left((XX^\dagger)^{\otimes 2} \mathcal{T}(X^\dagger X)^{\otimes 2} \right), O_{\text{ph}} ^{-1} ,\mathcal{T}^{-1}$ in the last line are all diagonal. Furthermore, the time evolution operator $\mathcal{T}$ factorizes in local diagonal operators acting on each vertex (see \eq{eq:ImaginaryTime}); all of these operators except those acting on the four vertices affected by the loop $(X^\dagger X)^{\otimes 2}$ commute with this loop operator and cancel with the respective inverse in $\mathcal{T}^{-1}$. The product of these three operators is thus local again. In total, this maps the four-body loop of $X$ operators to an at most 12-body diagonal operator. Alternatively, this diagonal operator can be obtained by a similar procedure as in the single-line case, by pulling the $X$ operators from the physical legs to virtual legs of the domain-wall tensor as
\begin{equation}
        \includegraphics[scale=0.425]{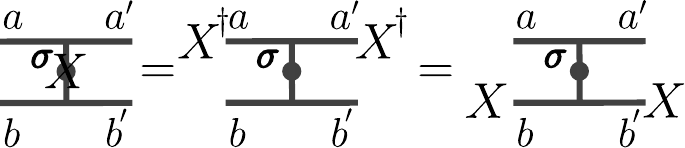}.
\end{equation}
From this, we see that every $X$ operator modifies the two $W$-matrices which are connected by its edge. For any product of $X$ operators, the expectation value immediately evaluates to zero if the closed-loop condition is violated for any of these matrices. Hence, to obtain a non-zero expectation value, every $W$-matrix has to be affected by an even number of $X$ operators which allows us to still write the $W$-matrix in the form of \eq{eq:WDLConstraint} with some modified tensor $A^\prime_{acd^\prime b^\prime}$; as expected, the four-edge loop $(XX^\dagger)^{\otimes 2}$ is the smallest operator with this property. The tensor $A^\prime_{acd^\prime b^\prime}$ is obtained from the original $A_{acd^\prime b^\prime}$ by shifting the affected indices, similar to how the $W$-matrix in the single-line case is affected by the $X$ operators. Then, the twelve-body operator derived in  \eq{eq:LoopMappingGeneral} can be constructed from the $A$-tensors analogously to \eq{eq:DiagVertexOperator}.

By repeated action of this mapping for the smallest loops, any bigger $X$ operator loop is also mapped to a diagonal operator which factorizes into a product of such 12-body local diagonal operators acting on all degrees of freedom around one plaquette. Together with the already diagonal $Z$ operators, from this every generalized Pauli string is again mapped to a diagonal operator which can be evaluated from a single-line state via the relation \eq{eq:DLtoSL}. If the $A$-tensor of the double-line state fulfills a normalization condition
\begin{equation}
    \sum_{b^\prime} \vert A_{acd^\prime b^\prime} \vert^2 = 1 , \; \forall a,c,d^\prime
    \label{eq:ANorm},
\end{equation}
the $W$-matrix of the associated single-line tensor is also normalized as given in \eq{eq:WNorm} and allows for an evaluation in a stochastic process. The update rule is then $p(\sigma \rho \vert \mu \nu) = \vert A_{acd^\prime b^\prime}\vert^2$, where the states in the circuit correspond to the domain-wall degrees of freedom of the double-line tensor defined as $\sigma = (a-b) \; \text{mod} \; N$ and likewise for the other three legs of the tensor. The mapping to a single-line state can also be seen as neglecting all complex phases in the $A$-tensor, $A_{acd^\prime b^\prime} \rightarrow \vert A_{acd^\prime b^\prime}\vert$, which leads to a deformed $\mathbb{Z}_N$ toric code which has a single-line description.

For both normalized single-line states and normalized double-line states which conserve the string-net virtual symmetries, general operators with a $k$-body support can be decomposed into at most $N^{2k}$ generalized Pauli strings. For each of these strings, the mapping to a diagonal operator can be performed, which allows for evaluation of this arbitrary operator. As the number of Pauli strings grows exponentially, this precludes efficient classical evaluation of a generic non-local operator; for finite support, the number of Pauli strings is also finite, which implies that generic operators may still be evaluated provided their support is not too large.

It should be noted that the mapping of an arbitrary operator to a diagonal operator works regardless of the normalization property. Thus, even for a general single-line $W$-matrix (or a general $A$-tensor respecting the virtual symmetries), every operator can be mapped to an expectation value in a classical statistical model governed by a Hamiltonian with local interactions between spins living on the vertices. The matrix $\vert W _v\vert^2$ contains the statistical weights from the interactions between the spin at vertex $v$ with its four neighbors~\cite{verstraete_criticality2006, liu_simulating2024}. These expectation values can then be evaluated classically by Monte Carlo updates in the space of classical configurations, provided the used algorithm equilibrates sufficiently quickly.

\section{E. Concrete definition of the path in Fig.~3}

In this section, we define the tensor network path indicated in \fig{fig:Skeletons} and further shown in \fig{fig:Path}. For each sub-path connecting two phases, the fixed-point states are identified by their virtual symmetries or, in the case of symmetry fractionalization, the action of a global symmetry on the local tensor. Starting from these endpoints, a symmetry-preserving path of isometric tensors is constructed, which connects both phases at a critical point which exhibits both symmetry patterns. On the level of the wavefunction, these paths illustrate how an analytically tractable phase transition can be achieved between different topological orders either by a change of the \emph{global} sign structure of the state or by modifying the set of allowed string branchings, or between different symmetry fractionalization patterns of the same topological order by changing the \emph{local} sign structure.

Along the entire path, we extract the correlation length $\xi = -1/\log{\vert\eta_2\vert}$ of the quantum state by diagonalizing the transfer matrix which is constructed from the parametrized $W$-matrices, where $\eta_2$ is the second-largest eigenvalue in magnitude~\cite{iqbal_study2018, xu_characterization2021}. At all critical points encountered along the path, $\xi$ diverges as some entries of the $W$-matrix vanish; this corresponds to the emergence of a $U(1)$ conservation law in the stochastic circuit, leading to hydrodynamics and algebraic correlations of equal-site operators.

\subsection{1. $\mathbb{Z}_2$ Toric Code to Double Semion}
The first transition we consider is between the $\mathbb{Z}_2 $ toric code and the double semion model~\cite{xu_tensor2018, liu_simulating2024}; these are the only possibilities for a $\mathbb{Z}_2$ abelian string-net model. Note that in the main text, we discuss two layers of these models; however, as they are independent of each other and deformed in the same way along the entire sub-path, it suffices to consider a single layer. The fixed point states of the toric code $\ket {\mathbb{Z}_2}$ and the double semion model $\ket {\text{DS}}$ are connected by a global unitary $U$ which is diagonal in $Z$ basis, $\ket {\text{DS}} = U \ket {\mathbb{Z}_2}$. They differ only in the sign structure of the possible loop configurations $c$: While for the toric code, all amplitudes can be chosen to be positive $\alpha_{c, \mathbb{Z}_2} = 1$, the double semion model always exhibits a non-trivial sign structure $\alpha_{c, \text{DS}} \neq 1$ for some $c$ which cannot be broken down to independently treating the qubits surrounding each vertex; the unitary $U$ allocates the relevant phase to each configuration $U \ket{c} = \alpha_{c, \text{DS}}\ket c$. This non-locality of the signs also necessitates the use of a double-line tensor network.

We consider a local tensor whose structure is given by \eq{eq:DoubleLine} and qubits as local edge degrees of freedom; the bond dimension of the tensor is thus $D = 4$. As the closed-loop constraint has to be respected, the $W$-matrix can be expressed more compactly in the 4-tensor $A_{acd^\prime b^\prime}$ as prescribed by \eq{eq:WDLConstraint}, where each index represents one of the four lines shown in \eq{eq:DoubleLine}. For the toric code fixed point, all entries can be chosen to be positive and equal weight as $A_{acd^\prime b^\prime}^{\mathbb{Z}_2} = \frac{1}{\sqrt{2}}$ for all $a,c, d^\prime, b^\prime$. The virtual symmetry which protects the topological order here is simply a spin flip applied to all legs:

\begin{equation}
    \includegraphics[scale=0.55]{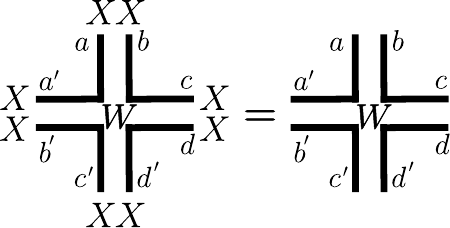}.
    \label{eq:VirtTC}
\end{equation}
The more involved sign structure of the double semion fixed point is reflected in the local tensor. Different prescriptions are possible~\cite{gu_tensor2009, levin_braiding2012, xu_tensor2018}; for our purposes, the relevant definition is $A^\text{DS}_{0011} = A^\text{DS}_{0110} = -\frac{1}{\sqrt{2}}$ and $A^\text{DS}_{acd^\prime b^\prime} = \frac{1}{\sqrt{2}}$ otherwise. This leads to a different kind of virtual symmetry with an additional diagonal operator:

\begin{equation}
    \includegraphics[scale=0.55]{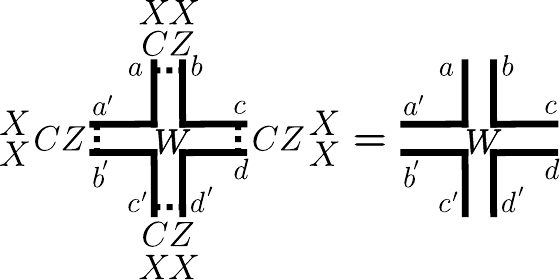},
    \label{eq:VirtDS}
\end{equation}
where the controlled-$Z$ gates act as $CZ \ket{11} = -\ket{11}$ and identity otherwise. The presence of the controlled-$Z$ gates leads to additional minus signs when concatenating virtual operators; this is indicative of the non-trivial self-statistics of the semionic vertex excitations of the double semion model, compared to the toric code.

We can connect these two fixed points by a path $A(g)$ which is parametrized by a single parameter $g \in [-1,1]$ as

\begin{align}
    & A_{0010}(g) =A_{0111}(g) = \frac{1}{\sqrt{1+\vert g \vert}}, \; A_{0011}(g) =A_{0110}(g) = \text{sign}(g) \sqrt{\frac{\vert g\vert}{1+\vert g \vert}}, \nonumber \\ & A_{1101}(g) =A_{1000}(g) = \frac{1}{\sqrt{1+\vert g \vert}}, \; A_{1100}(g) =A_{1001}(g) = \sqrt{\frac{\vert g\vert}{1+\vert g \vert}}
    \label{eq:PathTCDS}
\end{align}
and $A_{acd^\prime b^\prime}(g) = \frac{1}{\sqrt{2}}$ otherwise, with $A_{acd^\prime b^\prime}(g = 1) = A^{\mathbb{Z}_2}_{acd^\prime b^\prime}$ and $A_{acd^\prime b^\prime}(g = -1) = A^{\text{DS}}_{acd^\prime b^\prime}$. For $g \geq 0$, the system respects the toric code virtual symmetry of \eq{eq:VirtTC}, while for for $g \leq 0$, it respect the double semion virtual symmetry of \eq{eq:VirtDS}. Accordingly, at the transition point $g =0$ between the two topological orders, both symmetries are preserved.

The $A$-tensor remains normalized as given by \eq{eq:ANorm} for all configurations $a^\prime b^\prime c^\prime d^\prime $; for the full tensor, this implies an isometry property as

\begin{equation}
    \sum_{\rho,\sigma, a^\prime, b^\prime, c^\prime, d^\prime } T^{\sigma\rho}_{(ab)(cd)(a^\prime b^\prime )(c^\prime d^\prime )} (T^*)^{\sigma\rho}_{(a^{\prime\prime} b^{\prime\prime})(c^{\prime\prime}d^{\prime\prime})(a^\prime b^\prime )(c^\prime d^\prime )} = \delta_{a a^{{\prime\prime}}}\delta_{b b^{{\prime\prime}}}\delta_{cc^{{\prime\prime}}}\delta_{dd^{{\prime\prime}}}\delta_{bc}.
\end{equation}
Compared to the standard isometry condition of isoTNS, this equation features an additional constraint $\delta_{bc}$ which links the two outgoing legs; as a state built up from these tensors always fulfills this closed-loop condition, this does not affect the contraction properties or quantum circuit description of the normalized double-line tensors. It is further possible to lift the isometry in the subspace given by this projector to the whole target space; this extended tensor then fulfills the standard isoTNS condition~\cite{soejima_isometric2020}.

As discussed in the main text and Section B, the virtual symmetries fix the weights of virtual configurations which correspond to the same physical configurations. The double-line tensors can therefore be mapped to the local update rule of a stochastic process. Here, the rule of the stochastic brickwork automaton visualized in \fig{fig:Mapping} is given by $p(\mu \nu\vert \sigma \rho) = \vert A_{acd^\prime b^\prime} (g)\vert^2$, where the two sites in states $\sigma$ and $\rho$ are changed to $\mu$ and $\nu$, respectively. These states correspond to the physical domain walls of the double-line tensor as given by $\sigma = (a -b) \; \text{mod} \; 2$, $\rho = (c - d) \; \text{mod} \; 2$, $\mu = (a^\prime -b^\prime) \; \text{mod} \; 2$, $\nu = (c^\prime -d^\prime) \; \text{mod} \; 2$. This mapping is agnostic about the sign of $g$; this corresponds to a stochastic automaton with conservation of parity $P_{2} = (a+b) \; \text{mod} \; 2$, which features exclusively exponentially decaying correlations away from the point $g = 0$. At $g= 0$, two otherwise possible updates vanish as $p(10\vert01)=p(01\vert10)=0$; therefore, the local quantity $q = \sigma - \rho$ is conserved. Accordingly, the globally conserved quantity $Q = \sum_j q_j$ can be described using hydrodynamics and leads to algebraically decaying correlations in the time direction at the phase transition point. This transition can serve as a simple stand-in for phase transitions between string-net models with different topological order whose ground state wavefunctions differ in the complex phases of the amplitudes of the loop configurations; by tuning all amplitudes which are different between the two states to zero and having them re-appear with the new phase while staying in the plumbed isoTNS manifold, a direct transition can be realized.

\subsection{2. Two $\mathbb{Z}_2$ Toric Code layers to $\mathbb{Z}_4$ Toric Code}
The second transition on our parametrized path connects a pair of $\mathbb{Z}_2$ toric code states to a single $\mathbb{Z}_4$ toric code. The local Hilbert space on an edge has dimension $N = 4$; the two-level Hilbert spaces of the two $\mathbb{Z}_2$ toric codes can be mapped to this by $\ket {j_1} \ket {j_2} \rightarrow \ket{2j_1 +j_2}.$

For these phases, we can use a single-line description as given in \eq{eq:SingleLineSM} with $D = N = 4$. At the fixed point wavefunction $\ket{\mathbb{Z}_2^2}$, the pair of toric codes is completely decoupled; the state therefore is a positive equal-weight superposition of all configurations which respect the parity constraints of both layers at each vertex independently. The $W$-matrix of this state is given by $W_{(ab)(cd)}^{\mathbb{Z}_2^2} = \frac{1}{2}$ if $C_{1}(a,b,c,d) = \left( \frac{1}{2}\sum_{j =a,b,c,d} j - (j \; \text{mod} \; 2) \right) \; \text{mod} \; 2= 0$ and $C_{2}(a,b,c,d) = (a+b +c +d) \; \text{mod} \; 2= 0$, and $W_{(ab)(cd)}^{\mathbb{Z}_2^2} = 0$ otherwise. Introducing the operators $Z_1$ and $Z_2$ which act as $Z_1 \ket j = (-1)^{j \; \text{mod}\; 2} \ket j$ and $Z_2 \ket j = (-1)^{\frac{1}{2}(j - (j \; \text{mod} \; 2))} \ket j$ on a local degree of freedom, we can write the $\mathbb{Z}_2$ virtual symmetries of the two layers as

\begin{equation}
    \includegraphics[scale=0.55]{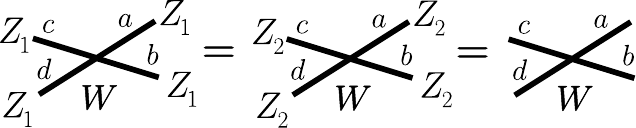}.
    \label{eq:VirtZ22}
\end{equation}
The $\mathbb{Z}_4$ toric code wavefunction $\ket{\mathbb{Z}_4}$ has a $W$-matrix with $W^{\mathbb{Z}_4}_{(ab)(cd)} = \frac{1}{2}$ if $C_{3}(a,b,c,d) = (a + b -c-d) \; \text{mod} \; 4= 0$ and $W^{\mathbb{Z}_4}_{(ab)(cd)} =0$ otherwise. The closed-loop constraint here is given in terms of the $N = 4 $ $Z$ operators acting on the $W$-matrix as

\begin{equation}
    \includegraphics[scale=0.55]{ZConstraint.pdf}.
    \label{eq:VirtZ4}
\end{equation}
For both fixed point $W$-matrices, there are certain configurations which violate the virtual symmetry of the other fixed point. A possible way to continuously deform them is to tune the weight of these entries to zero, connecting them at a critical point where the virtual symmetries of both phases are preserved. For the two-layer $\mathbb{Z}_2$ phase, a possible parametrization which respects the virtual symmetry (\ref{eq:VirtZ22}) of the fixed point is $W^1(g)$ with $g \in [0,1]$ defined as

\begin{flalign}
     W^1_{(ab)(cd)}(g) =\frac{1}{\sqrt{2+2g}} \; \text{if} \; &C_{1}(a,b,c,d) = C_{2}(a,b,c,d) = C_{3}(a,b,c,d) =0, && \nonumber\\ &C_{1}(a +1\; \text{mod} \; N,b+1 \; \text{mod} \; N,c,d) =C_{2}(a +1\; \text{mod} \; N,b+1 \; \text{mod} \; N,c,d) = 0 ,\nonumber &&\\  W^1_{(ab)(cd)}(g) =\sqrt{\frac{g}{2+2 g }} \; \text{if} \; & C_{1}(a,b,c,d) = C_{2}(a,b,c,d) =0, C_{3}(a ,b,c,d) \neq 0, \nonumber&& \\ W^1_{(ab)(cd)}(g) =\frac{1}{2} \;\;\;\;\;\;\;\;\;\;\;\;\; \text{if} \; & C_{1}(a,b,c,d) = C_{2}(a,b,c,d) = C_{3}(a,b,c,d) =0,\nonumber&& \\ & C_{1}(a +1\; \text{mod} \; N,b+1 \; \text{mod} \; N,c,d) =C_{2}(a +1\; \text{mod} \; N,b+1 \; \text{mod} \; N,c,d) \neq 0&&
\end{flalign}
and $W^1_{(ab)(cd)}(g) = 0$ otherwise. $W^1_{(ab)(cd)}(g = 1) = W_{(ab)(cd)}^{\mathbb{Z}_2^2}$, while for $g < 1$ the two layers are coupled to each other. The two distinct prescriptions for configurations which conserve all virtual symmetries is necessary to ensure the $W$-matrix remains normalized along the entire path. The associated stochastic update rule $p(cd \vert ab) = \vert W^1_{(ab)(cd)} \vert^2$ conserves the two parity-like quantities $P_{2,1} = \frac{1}{2} \Bigl(a - (a \; \text{mod} \; 2) + b - (b \; \text{mod} \; 2)\Bigr)$ and $P_{2,2} = (a+b) \; \text{mod} \; 2$.

With the same logic, the $\mathbb{Z}_4$ fixed point can be connected via another path $W^2(g)$ with $g \in [0,1]$ defined as

\begin{flalign}
     W^2_{(ab)(cd)}(g) =\frac{1}{\sqrt{2+2g}} \; \text{if} \; &C_{1}(a,b,c,d) = C_{2}(a,b,c,d) = C_{3}(a,b,c,d) =0, && \nonumber\\ &C_{1}(a +1\; \text{mod} \; N,b-1 \; \text{mod} \; N,c,d) =C_{2}(a +1\; \text{mod} \; N,b-1 \; \text{mod} \; N,c,d) \neq 0 ,\nonumber &&\\  W^2_{(ab)(cd)}(g) =\sqrt{\frac{g}{2+2 g }} \; \text{if}  \;& C_{1}(a,b,c,d) = C_{2}(a,b,c,d) \neq 0, C_{3}(a ,b,c,d) = 0, \nonumber&& \\ W^2_{(ab)(cd)}(g) =\frac{1}{2} \;\;\;\;\;\;\;\;\;\;\;\;\; \text{if} \;& C_{1}(a,b,c,d) = C_{2}(a,b,c,d) = C_{3}(a,b,c,d) =0,\nonumber&& \\ & C_{1}(a +1\; \text{mod} \; N,b-1 \; \text{mod} \; N,c,d) =C_{2}(a +1\; \text{mod} \; N,b-1 \; \text{mod} \; N,c,d) = 0&&
\end{flalign}
and $W^2_{(ab)(cd)}(g) = 0$ otherwise, with $W^2_{(ab)(cd)}(g = 1) = W_{(ab)(cd)}^{\mathbb{Z}_4}$. This path respects the virtual symmetry of \eq{eq:VirtZ4}, which corresponds to conservation of the quantity $P_4 = (a + b) \; \text{mod} \; 4$ in the stochastic circuit.

The two paths meet at the endpoint where the weights of the symmetry-violating configurations of the other phase vanish, $W^1_{(ab)(cd)}(g = 0) = W^2_{(ab)(cd)}(g = 0)$. This critical point is invariant under both types of virtual symmetries. The stochastic process with $p(cd\vert ab) = \vert W_{(ab)(cd)}\vert^2$ locally conserves the quantity $q_j = a \; \text{mod} \; 2 + b \; \text{mod} \; 2$, leading to a $U(1)$ conservation law for the global quantity $Q = \sum_j q_j$ and to a diverging correlation length with algebraically decaying correlations in the time-like direction. This is the simplest case of coupling $k$ copies of a $\mathbb{Z}_N$ toric code state to a $\mathbb{Z}_{N^k}$ toric code; at the critical point the allowed types of loop coverings change, prompting a change of the topological order at the the phase transition. This can be contrasted with the class of transitions discussed in the preceding section, where the topological phase transition is induced by changing the complex phases of a fixed set of configurations.

\subsection{3. Two $\mathbb{Z}_4$ Toric Code States with different symmetry fractionalization}
The transitions in the two preceding sections concern phases with different topological order, i.e. where the anyons differ in their braiding or fusion statistics. However, in the presence of additional global symmetries, topological phases can further be distinguished by how their anyonic excitations transform under these symmetries~\cite{chen_symmetry_2017}. These topological states with global symmetries are called \emph{symmetry-enriched topological} (SET) phases. For a state with some distant excitations, the symmetry \emph{fractionalizes} on the excitations; as the ground state is invariant under the symmetry operator $O$, the application of a global symmetry operator can be reduced to local unitary operators $U(O)$ around the excitations. If these operators form a non-trivial representation of the symmetry group, the anyons feature non-trivial symmetry fractionalization, adding further structure to the excitation content of the state.

Phases with different symmetry fractionalization schemes constitute different orders only as long as the symmetry remains unbroken; otherwise, they can be connected by an adiabatic path. For tensor network states, a sufficient condition that a state is symmetric under some on-site symmetry is the ``pulling-through" condition~\cite{williamson_symmetry2017}: This requires that the application of the local symmetry operator on the physical degrees of freedom of a tensor allows us to pull through some matrix product operator from one pair of virtual legs to the other; a global application of the symmetry then leaves the state invariant.

In this section, we describe an isoTNS single-line path which interpolates between two $\mathbb{Z}_N$ toric code fixed points for even $N$ which feature different fractionalization patterns under a global anti-unitary $\mathbb{Z}_2^T$ symmetry given by $\{ 1, \mathcal{K} X^{N/2} \}$, where $\mathcal{K}$ is complex conjugation; in the main text, the path is specified to $N = 4$. Such a symmetry allows for two representations acting locally on the $\mathbb{Z}_N$ anyons; it can either act trivially on these excitations, if the representation is linear as $U(\mathcal{K} X^{N/2})U^*(\mathcal{K} X^{N/2}) = 1$, or non-trivially for a projective representation $U(\mathcal{K} X^{N/2})U^*(\mathcal{K} X^{N/2}) = -1$. In the second case, the excitations exhibit Kramer's degeneracy and are thus doubly degenerate under the symmetry~\cite{haller_quantum_2023}.

An important advantage of the plumbing states considered here is that the local action of the symmetry operators can be pulled to virtual level as given by \eq{eq:Plumbing}; we will use this relation to construct the fixed points of the phase exhibiting non-trivial symmetry fractionalization. The fixed points $\ket{\mathbb{Z}_N}$ given by a single-line tensor network and a strictly positive $W$-matrix with $W^{\mathbb{Z}_N}_{(ab)(cd)} = \frac{1}{\sqrt{N}}$ for $(a+b) - (c+d) = 0 \; \text{mod} \; N$ and zero otherwise exhibit trivial symmetry fractionalization on the vertex excitations $e$ which are created by strings of $X$ operators; this can be seen by the local action of the symmetry operator pulled to the virtual level:

\begin{equation}
    \includegraphics[scale=0.55]{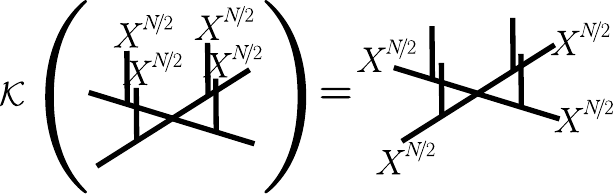}
    \label{eq:FracTriv}
\end{equation}
As $X^N = 1$, the representation on the vertex excitations is linear. The fixed points $\ket{\mathbb{Z}_{N,f}}$ with non-trivial symmetry fractionalization are given by a $W$-matrix with a different sign structure $W^{\mathbb{Z}_{N,f}}_{(ab)(cd)} = \frac{\text{sign(a,b,c,d)}}{\sqrt{N}}$ for $(a+b) - (c+d) = 0 \; \text{mod} \; N$ and zero otherwise. If $N = 4k$ is divisible by four, $\text{sign(a,b,c,d)} = -1$ if $a + N b \leq N^2/2$ and an odd number of $a,b,c,d$ are odd numbers $> N/2$; if $N = 4k +2$, $\text{sign(a,b,c,d)} = -1$ if $a + N b \leq N^2/2$ and either $a,b$ are both even and $c,d$ both odd or vice versa. Otherwise, $\text{sign(a,b,c,d)} = 1$. This convoluted definition leads to a different matrix product operator representing the local symmetry action, which is

\begin{equation}
    \includegraphics[scale=0.55]{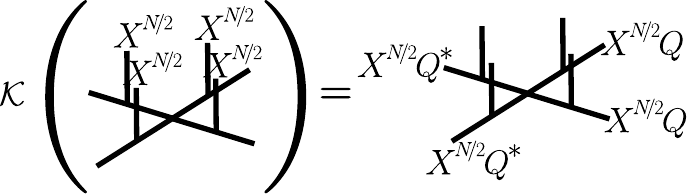}.
    \label{eq:FracNonTriv}
\end{equation}
$Q$ is a diagonal operator with $Q_{jj} = -1$ if $j$ is odd and $> N/2$ for $N = 4k$ and $Q_{jj} = i$ if $j$ is odd for $N = 4k +2$, otherwise $Q_{jj} = 1$. Applying the symmetry twice thus leads to a virtual operator $X^{N/2}Q^*X^{N/2}Q \propto Z^{N/2}$. Thus, locally at a single vertex applying the symmetry operator twice is equivalent to the vertex operator $A_v^{N/2} $; this returns $-1$ for an odd number of vertex excitations $e$. The symmetry thus fractionalizes non-trivially on these anyons.

As an aside, we note that the global symmetry $\mathcal{K} X^{N/2}$ also has a non-trivial action on the plaquette excitations $m$ which are created by strings of $Z$ operators; as $(Z^k)^* = Z^{N-k}$, the anyons $m^k$ are permuted with their respective anti-particles $m^{N-k}$. It is in fact a general property that an anti-unitary $\mathbb{Z}_2^T$ symmetry has to permute some anyons for $N > 2$~\cite{essin_classifying2013}.

The fixed points can be connected by a symmetry-preserving single-line tensor path. We want the sign structure to change when crossing from trivial fractionalization to non-trivial fractionalization. Introducing the parameter $g \in [-1,1]$, we first define the prefactor function $P_{abcd}(g)$ with $P_{abcd}(g) =1$ if $g \geq 0$ and $P_{abcd}(g) =\text{sign(a,b,c,d)}$ if $g <0$. In the case $N = 4k$, the path is parametrized in the following way: For $(a+b) - (c+d) = 0 \; \text{mod} \; N$, $W_{(ab)(cd)}(g) =\frac{P_{abcd}(g)}{\sqrt{1+\vert g\vert}} \frac{\sqrt{2\vert g\vert}}{\sqrt{N}}$ if an odd number of $a,b,c,d$ are odd numbers $> N/2$ \emph{or} if either $a,b$ are both even and $c,d$ both odd or vice versa; otherwise $W_{(ab)(cd)}(g) =\frac{1}{\sqrt{1+\vert g\vert}} \frac{\sqrt{2}}{\sqrt{N}}$. All other entries are zero. In the case $N = 4k +2$, the parametrization is similar: For $(a+b) - (c+d) = 0 \; \text{mod} \; N$, $W_{(ab)(cd)}(g) =\frac{P_{abcd}(g)}{\sqrt{1+\vert g\vert}} \frac{\sqrt{2\vert g\vert}}{\sqrt{N}}$ if either $a,b$ are both even and $c,d$ both odd or vice versa; $W_{(ab)(cd)}(g) =\frac{1}{\sqrt{1+\vert g\vert}} \frac{\sqrt{2}}{\sqrt{N}}$ if $a,b,c,d$ are all even or all odd; otherwise $W_{(ab)(cd)}(g) = \frac{1}{\sqrt{N}}$. Other entries are zero. Along the entire path, the $W$ matrix respects the virtual symmetry of \eq{eq:ZVirtSymm}, corresponding to the conservation of the parity-like quantity $P_N = (a+b) \; \text{mod} \; N$ in the local update rule.

For $g \geq 0$, the virtual operator corresponding to the local symmetry operator is given by \eq{eq:FracTriv}, while for $g \leq 0$ it is given by \eq{eq:FracNonTriv}; for $g = 0$, both operators are valid as the configurations for which they differ have vanishing weight. The symmetry fractionalization pattern on the vertex operator can be probed using a membrane order parameter $\langle O_{e} \rangle$. This order parameter is defined on an infinite cylinder with circumference $L$. Introducing the minimally entangled state $\ket {\psi_e}$ corresponding to the vertex excitation $e$, it is defined as the normalized expectation value of partial application of the unitary part of the symmetry $U_{\text{part}}$ in the thermodynamic limit as

\begin{equation}
    \langle O_{e} \rangle = \lim_{L \rightarrow \infty} \vert \bra {\psi_e}U_{\text{part}}\ket {\psi_e}\vert^{1/L}.
\end{equation}
Here, $U_{\text{part}} = \prod_i X_i^{N/2}$ acts only on a subsystem in the bulk. As to avoid generating further excitations, it should be composed as $U_{\text{part}} = \prod_p B_p^{N/2}$, where $B_p = (XX^\dagger)^{\otimes 2}$ is the toric code plaquette operator. This expectation value vanishes for non-trivial symmetry fractionalization and takes a finite value for trivial symmetry fractionalization; in \fig{fig:Path} in the main text, we also show its value in the relevant phases for $N = 4$ from corner transfer matrix renormalization group numerics~\cite{nishino_corner1996, haller_quantum_2023, liu_simulating2024, boesl_quantum2025}. (The cases with $\langle O_{e} \rangle = 1$ can be evaluated exactly.)

It should be noted that the states described by $W_{(ab)(cd)}(g)$ and $W_{(ab)(cd)}(-g)$ are connected by a finite-depth unitary circuit; this has to be contrasted with the states in subsection 1, which can be connected only by a global unitary. In contrast, the states here differ only in symmetry fractionalization patterns, not in the actual topological order. The sign structure is thus different only by local changes which can be expressed in a single-line tensor. The associated local unitary breaks the $\mathbb{Z}_2^T$ symmetry, as is necessary to reach a different SET order.

As in the preceding cases, the transition point given by $W_{(ab)(cd)}(g = 0)$ exhibits a conservation law in the associated stochastic circuit which leads to a diverging correlation length. Here, the local quantity $q_j = a \; \text{mod} \; 2 + b \; \text{mod} \; 2$ is conserved; this is the same conservation law as discussed in the previous section. The stochastic circuit at the critical point between the two SET orders has fewer allowed transitions; in fact, some updates which conserve this local quantity have zero probability, while for the critical state between the two different topological orders in subsection 2, all valid updates have finite probability. This allows us to tune between the two critical points on a path which exhibits the conservation law along the entire line and thus features the same critical correlations; this is indicated in \fig{fig:Skeletons} by the red line connecting the two critical points.

More generally speaking, for $N = 4k$, the path described here also changes the weight of some configurations which are not affected by the difference in the virtual operators; it would also be possible to leave them invariant. The associated circuit then does not exhibit critical correlations at the phase transition point $g = 0$ as no conservation law is present in the dynamical process; nevertheless, a gap closes at this point as another state becomes degenerate with the ground state manifold. The phases are therefore still well-distinguished as expected for a symmetry-preserving path.

\section{F. Iso-TNS from higher multipole-conserving processes}
The algebraic decay of time-like correlations at all critical points on the path discussed in the main text and Section E results from the emergence of some conserved quantity $Q = \sum_j q_j$, where $j$ runs over gates in one update step of the stochastic circuit. This allows for a hydrodynamic description, which leads to a diffusion equation for the local density $q_j$. As these local quantities correspond to diagonal operators acting on the isoTNS wavefunction, some correlator built from $Z$ operators exhibits diffusive decay $\langle Z^k(x,y) (Z^\dagger)^k(x,y+r) \sim r^{-1/2}$, where the coordinate $y$ represents the direction corresponding to time $t$ in the stochastic automaton. The diffusive dynamical exponent $z = 2$ can be modified in the presence of additional conservation laws to subdiffusive values $ z > 2$. In particular, if multipole moments of the charge $Q = Q^{(0)}$ of the form $Q^{(n)} = \sum_j j^n q_j$ are conserved, the dynamics are slowed down. Concretely, a stochastic automaton which conserves the first $m$ multipole moments $Q^{(n)}$ with $n \leq m$ while remaining ergodic is governed by a dynamical exponent $z = 2(m+1)$~\cite{gromov_fracton2020, feldmeier_anomalous2020}. From the mapping between these automata and single-line isoTNS, this implies there is a class of isoTNS with critical correlations in one direction of the form

\begin{equation}
    \langle Z^k(x,y) (Z^\dagger)^k(x,y+r) \rangle \sim r^{-1/2(m+1)}.
\end{equation}
In this section, we explicitly construct such an isoTNS wavefunction for the case of dipole moment conservation $m = 1$. We also discuss how this critical wavefunction can be connected to a toric code fixed point via a wavefunction which passes through a sequence of critical $m =0$ wavefunctions, in a logic similar to the paths connecting distinct topological phases.

\begin{figure*}
    \includegraphics[width=\textwidth]{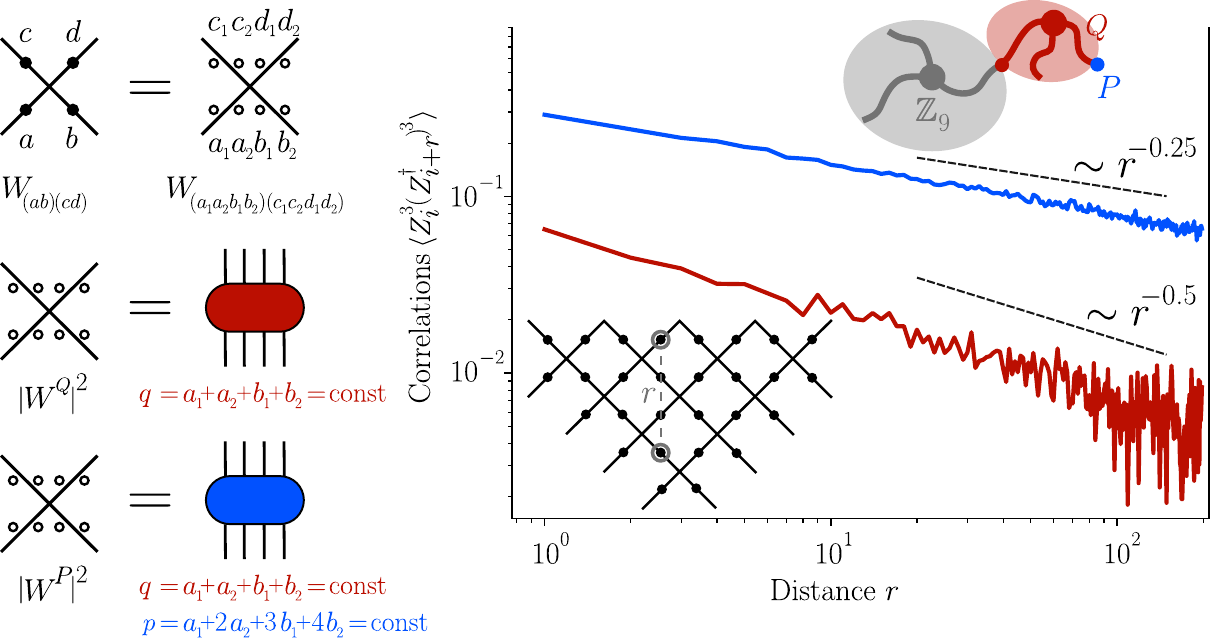}
    \caption{\label{fig:Dipole}
        \textbf{Dipole isoTNS.} (Left column) By splitting up a $9$-level edge degree of freedom into two qutrits, the corresponding normalized $W$-matrix can be mapped to a 4-local gate of a stochastic process. The weights can be chosen in such a way as to conserve a global $U(1)$ charge $Q = \sum_j q_j$ in the time evolution (red) or the charge $Q$ alongside the associated dipole moment $P = \sum_j jq_j$ (blue). (Right column) These conservation laws lead to diffusive equal-position correlations $\sim r^{-1/2}$ for a conserved charge and subdiffusive correlations $\sim r^{-1/4}$ for conserved charge and dipole moment which are reflected in the correlations of the respective isoTNS. The numerical results are evaluated starting from an open boundary of two edges meeting at a corner (lower left inset). The charge-conserving critical reference state given by $W^Q$ has isoTNS deformations of its own consisting of paths which respect charge conservation of the stochastic process; it can be continuously connected to a topological phase such as the $\mathbb{Z}_9$ toric code or to the dipole critical state by passing through a path on the $N = 9$ isoTNS skeleton (top right inset).
    }
\end{figure*}

The two-local gates we discuss in the main text cannot realize non-trivial dipole moment conserving dynamics, as any local move besides leaving the configuration invariant changes the dipole moment. The gates in question thus need to have higher support. To translate this into a $W$-matrix, we take a state with a local edge Hilbert space of dimension $N = 9$ and rewrite it as seen depicted in the top right of \fig{fig:Dipole}; each $9$-level qudit is represented by a pair of $3$-level qutrits, with the states being identified as $\ket{j_1}\ket{j_2} \rightarrow \ket {3j_1 + j_2}$. The squared norm of a normalized $W$-matrix can then be interpreted as a classical update rule acting on four $3$-level sites, the probability of one step being $p(c_1c_2d_1d_2\vert a_1a_2b_1b_2) = 
\vert W_{(a_1a_2b_1b_2)(c_1c_2d_1d_2)}\vert^2$. If we denote as $N^{(0)}_{a_1a_2b_1b_2}$ the number of allowed configurations for $c_1,c_2,d_1,d_2$ such that $a_1+a_2+b_1+b_2=c_1+c_2+d_1+d_2$, the matrix with entries $W^{Q}_{(a_1a_2b_1b_2)(c_1c_2d_1d_2)} = \frac{1}{\sqrt{N^{(0)}_{a_1a_2b_1b_2}}}$ if $a_1+a_2+b_1+b_2=c_1+c_2+d_1+d_2$ and $0$ otherwise conserves the local charge $q = a_1+a_2+b_1+b_2$. Thus the global quantity $Q = Q^{(0)}= \sum_j q_j$ is conserved. As $Z^3 \ket{j} = e^{2\pi ij_2/3} \ket{j}$, we would thus expect $\langle Z^3(x,y) (Z^\dagger)^3(x,y+r) \rangle \sim r^{-1/2}$.

However, we may also impose the additional constraint of dipole moment conservation: If $N^{(1)}_{a_1a_2b_1b_2}$ is the number of configurations $c_1,c_2,d_1,d_2$ such that $a_1+a_2+b_1+b_2=c_1+c_2+d_1+d_2$ and $a_1+2a_2+3b_1+4b_2=c_1+2c_2+3d_1+4d_2$, the matrix $W^{P}_{(a_1a_2b_1b_2)(c_1c_2d_1d_2)} = \frac{1}{\sqrt{N^{(1)}_{a_1a_2b_1b_2}}}$ if these conditions are fulfilled and $0$ otherwise corresponds to a circuit which conserves not only the global charge $Q$, but also its associated dipole moment $P = Q^{(1)} = \sum_j jq_j$. This state thus features more slowly decaying correlations $\langle Z^3(x,y) (Z^\dagger)^3(x,y+r) \rangle \sim r^{-1/4}$. In \fig{fig:Dipole}, we show numerical results for these two states; the two critical exponents can clearly be distinguished. The states are defined with open boundaries; the boundary consists of two edges which meet at a corner which serves as the first time step $t = 0$ of the stochastic process (see inset). We choose a fully disordered boundary $\ket{\psi_0} = \ket{++ \cdots +}$, where $\ket{+} = \frac{1}{3} \sum_{j=0}^9 \ket j$; the correlations can then be evaluated by averaging over the trajectories of the stochastic process.

This logic can be applied to construct stochastic circuits with any number of conserved higher multipoles; the necessary local Hilbert space dimension increases for each additional conserved multipole moment for two reasons: First, more classical sites have to be placed at each edge to allow for updates which change the configuration while respecting all conservation laws; second, to ensure the process remains ergodic and is not frozen, the number of possible states at each classical site has to increase as well. In the present example of dipole moment conservation, a 4-local update of sites with local Hilbert space dimension $2$ would not produce subdiffusion. Interpreting the time evolution of the circuit as the motion of classical particles, most configurations would remain stuck as particles are not allowed to pass each other, thereby freezing out the dynamics. Increasing the local dimension to $3$ states restores ergodicity as particles do not obstruct each other anymore, preventing further fragmentation of the charge and dipole sectors~\cite{sala_ergodicity2020}.

It is furthermore possible to connect these states to the fixed points of topological phases. If we consider the fixed point $\ket{\mathbb{Z}_9}$ of the $N = 9$ toric code phase given by the single-line tensor network with matrix $W^{\mathbb{Z}_9}_{(ab)(cd)} = \frac{1}{3}$ if $( a + b) - (c+d) = 0 \; \text{mod} \; 9$ and $0$ otherwise, we can construct a path towards $W^P$ in a piecewise fashion: Defining $N^{(\mathbb{Z}_9,0)}_{a_1a_2b_1b_2}$ as the number of configurations $c_1, c_2, d_1, d_2$ which fulfill both the $\mathbb{Z}_9$ parity constraint and particle number conservation, the first sub-path is given by $W^1_{(a_1a_2b_1b_2)(c_1c_2d_1d_2)}(g)$ with $g \in [0,1]$ as

\begin{flalign}
     W^1_{(a_1a_2b_1b_2)(c_1c_2d_1d_2)}(g) =\frac{1}{\sqrt{N^{(\mathbb{Z}_9,0)}_{a_1a_2b_1b_2}+(9-N^{(\mathbb{Z}_9,0)}_{a_1a_2b_1b_2})g}} \; \text{if} \; &( a_1 + a_2+ b_1 + b_2) - (c_1 + c_2+d_1 + d_2) = 0 \; \text{mod} \; 9,&& \nonumber\\ &a_1+a_2+b_1+b_2=c_1+c_2+d_1+d_2,\nonumber &&\\  W^1_{(a_1a_2b_1b_2)(c_1c_2d_1d_2)}(g) =\sqrt{\frac{g}{N^{(\mathbb{Z}_9,0)}_{a_1a_2b_1b_2}+(9-N^{(\mathbb{Z}_9,0)}_{a_1a_2b_1b_2})g}} \; \text{if} \; & ( a_1 + a_2+ b_1 + b_2) - (c_1 + c_2+d_1 + d_2) = 0 \; \text{mod} \nonumber\\ &a_1+a_2+b_1+b_2 \neq c_1+c_2+d_1+d_2
\end{flalign}
and $0$ otherwise. This parametrization interpolates from $W^1(g= 1)=W^{\mathbb{Z}_9}$ to a state $W^1(g= 0)$ with particle number conservation and thus critical correlations. Defining a second part as $W^2_{(a_1a_2b_1b_2)(c_1c_2d_1d_2)}(g)$ with $g \in [0,1]$ as

\begin{flalign}
     W^2_{(a_1a_2b_1b_2)(c_1c_2d_1d_2)}(g) =\frac{1}{\sqrt{N^{(\mathbb{Z}_9,0)}_{a_1a_2b_1b_2}+(N^{(0)}_{a_1a_2b_1b_2}-N^{(\mathbb{Z}_9,0)}_{a_1a_2b_1b_2})g}} \; \text{if} \; &( a_1 + a_2+ b_1 + b_2) - (c_1 + c_2+d_1 + d_2) = 0 \; \text{mod} \; 9,&& \nonumber\\ &a_1+a_2+b_1+b_2=c_1+c_2+d_1+d_2,\nonumber &&\\  W^2_{(a_1a_2b_1b_2)(c_1c_2d_1d_2)}(g) =\sqrt{\frac{g}{N^{(\mathbb{Z}_9,0)}_{a_1a_2b_1b_2}+(N^{(0)}_{a_1a_2b_1b_2}-N^{(\mathbb{Z}_9,0)}_{a_1a_2b_1b_2})g}} \; \text{if} \; & ( a_1 + a_2+ b_1 + b_2) - (c_1 + c_2+d_1 + d_2) \neq 0 \; \text{mod}\; 9\nonumber\\ &a_1+a_2+b_1+b_2 = c_1+c_2+d_1+d_2
\end{flalign}
and $0$ otherwise, this connects the state $W^2(g= 0) = W^1(g= 0)$ to the charge-conserving reference state $W^2(g= 1) = W^Q$. This state is then continuously connected to the dipole moment conserving state via a third sub-path $W^3_{(a_1a_2b_1b_2)(c_1c_2d_1d_2)}(g)$ with $g \in [0,1]$:

\begin{flalign}
     W^3_{(a_1a_2b_1b_2)(c_1c_2d_1d_2)}(g) =\frac{1}{\sqrt{N^{(1)}_{a_1a_2b_1b_2}+(N^{(0)}_{a_1a_2b_1b_2}-N^{(1)}_{a_1a_2b_1b_2})g}} \; \text{if} \; &a_1+2a_2+3b_1+4b_2=c_1+2c_2+3d_1+4d_2,&& \nonumber\\ &a_1+a_2+b_1+b_2=c_1+c_2+d_1+d_2,\nonumber &&\\  W^3_{(a_1a_2b_1b_2)(c_1c_2d_1d_2)}(g) =\sqrt{\frac{g}{N^{(1)}_{a_1a_2b_1b_2}+(N^{(0)}_{a_1a_2b_1b_2}-N^{(1)}_{a_1a_2b_1b_2})g}} \; \text{if} \; & a_1+2a_2+3b_1+4b_2\neq c_1+2c_2+3d_1+4d_2 \nonumber\\ &a_1+a_2+b_1+b_2 = c_1+c_2+d_1+d_2
\end{flalign}
and $0$ otherwise; here $W^3(g= 0) = W^Q$ and $W^3(g= 1) = W^P$. In total, this path continuously connects the topological phase with a gapless state with different criticality compared to those in the main text.

The deformations of the state constructed from $W^Q$ follow a similar logic as for the topological phases discussed in the main text. They are thus also part of the respective isoTNS skeleton, as indicated in the top left inset of \fig{fig:Dipole}: By respecting the charge conservation symmetry and the isometry condition, paths can be identified which leave the qualitative correlation structure unchanged. By tuning to endpoints which respect other conditions we can connect this ``critical" part of the isoTNS skeleton to gapped topological phases in the case of the $\mathbb{Z}_9$ constraint, or to states with modified criticality in the case of dipole conservation.

\end{document}